\documentclass{aa}
\usepackage[T1]{fontenc}
\DeclareRobustCommand{\VAN}[3]{#2}
\let\VANthebibliography\thebibliography
\def\thebibliography{\DeclareRobustCommand{\VAN}[3]{##3}\VANthebibliography}

\usepackage{graphicx}
\usepackage{txfonts}
\usepackage{hyperref}
\usepackage{graphicx}	
\usepackage{amsmath}	
\usepackage{tabularx}
\usepackage{booktabs}
\usepackage{multirow}
\usepackage{newtxtext,newtxmath}
\usepackage{comment}
\usepackage{lscape}
\usepackage{xcolor}
 
\usepackage[flushleft]{threeparttable}
\begin{document}

   \title{On the nature of the ultraluminous X-ray source Holmberg II X-1}

   \author{F. Barra\inst{1,2}\fnmsep\thanks{francesco.barra@unipa.it},
C. Pinto\inst{2},
M. Middleton\inst{3},
T. Di Salvo\inst{1},
D. J. Walton\inst{4},
A. Gúrpide\inst{3}
and
T. P. Roberts\inst{5}
}

\institute{
Universit\`a degli Studi di Palermo, Dipartimento di Fisica e Chimica, via Archirafi 36, I-90123 Palermo, Italy 
\and
INAF/IASF Palermo, via Ugo La Malfa 153, I-90146 Palermo, Italy
\and
Department of Physics \& Astronomy. University of Southampton, Southampton SO17 1BJ, UK
\and
Centre for Astrophysics Research, University of Hertfordshire, College Lane, Hatfield AL10 9AB, UK 
\and
Centre for Extragalactic Astronomy \& Dept of Physics, Durham University, South Road, Durham DH1 3LE, UK 
}

   \date{Received XXX; accepted XXX}

 
  \abstract
{We present a comprehensive spectral analysis of the ultraluminous X-ray source Holmberg II X-1 using broadband and high-resolution X-ray spectra taken with the XMM-\textit{Newton} satellite over a period of 19 years \textcolor{black}{benefiting from a recent campaign}. We tested several models for the broadband spectra among which a double thermal component provided a reasonable description for the continuum \textcolor{black}{between 0.3-10 keV} and enabled us to constrain the properties of the accretion disc. 
The Luminosity-Temperature trends of the inner and outer disc components \textcolor{black}{broadly agree with the expectations for a thin disc}, although the exact values of the slopes are slightly sensitive to the adopted model. However, all tested models show L-T trends which deviate from a \textcolor{black}{power law} above a bolometric luminosity of about $5 \times 10^{39} \rm erg/s$, \textcolor{black}{particularly for the hot thermal component associated to the inner accretion flow}. Assuming that such deviations are due to the \textcolor{black}{accretion rate exceeding its Eddington limit or, most likely, the super-critical rate, a compact object with a mass 16-36 M$_{\odot}$, i.e. a stellar-mass black hole, is inferred}. The time-averaged (2021) high resolution spectra present narrow emission lines at 1 keV primarily from Ne\,{\scriptsize IX-X} and a very strong at 0.5 keV from N\,{\scriptsize VII}, which indicate Ne-N-rich gas with non-Solar abundances. This favours a nitrogen-rich donor star, such as a blue/red supergiant, which has escaped from its native stellar cluster  \textcolor{black}{characterised by} a low-metallicity environment.}

\authorrunning{F. Barra et al. }
\titlerunning{On the nature of the ultraluminous X-ray source Holmberg II X-1}


   \maketitle
%

\section{Introduction}
\label{Sec:Introduction}
Ultraluminous X-ray sources (ULXs) are off-nucleus, point-like, extragalactic sources with X-ray luminosities above $L_{\rm X} > 10^{39}  \rm  erg/s$ (for recent reviews, see \citealt{Pinto_2023,King2023NewAR}), brighter than the Eddington limit for a black hole of 10 {M\textsubscript{\(\odot\)}}, resulting from accretion onto a compact object. The luminosity of a ULX can reach $10^{41} \rm erg/s$ in the X-ray band (0.3-10 keV) alone leading to many conjectures on the nature of the compact object. ULXs, in fact, were theorized as BHs,
with a mass greater than 10 and up to $10^{5}$ {M\textsubscript{\(\odot\)}} (intermediate mass black hole, IMBH, \citealt{Miller_2004}) with ESO 243-49 HLX-1  as the best candidate (\citealt{Webb2012}). But the discovery of X-ray pulsations in a fraction of the population of ULXs revealed that at least some ULXs are powered by neutron stars (NSs). Notable examples are M82 X-2 (\citealt{Bachetti_2014}), NGC 5907 ULX-1 (\citealt{Fuerst_2016}) and NGC 7793 P13 (\citealt{Israel_2017a}). \textcolor{black}{The number of persistent ULXs which exhibit pulsations is currently 6 with another 6 transient pulsars whose X-ray luminosity exceeded $10^{39}  \rm  erg/s$ for a short period (see Table 2 in \citealt{King2023NewAR}).} Given that about 30 systems have sufficiently good statistics to detect pulsations, the fraction of NS-powered ULXs might be around 30\% or above (\citealt{Rodriguez_2020}, see also \citealt{Middleton_King2017,King2020}).

ULX spectra are marked out by a curvature in the range 2-10 keV and a soft excess below 2 keV (the `ultraluminous state', \citealt{Gladstone_2009}) and can typically be classified into three different regimes, according to their spectral slope  $\Gamma$ in the 0.3-10 keV energy range: the soft ultraluminous (SUL) regime for $\Gamma > 2$, hard ultraluminous regime  (HUL) for $\Gamma<2$ and broadened disc (BD), the latter where the X-ray spectrum presents a single peak and is dominated by a blackbody-like component in the 2–5 keV band  (\citealt{Sutton_2013}). Several ULXs sometimes show spectra switching between these regimes (\citealt{Middleton_2015a}, \citealt{Gurpide_2021b}).
If the ULX spectrum is dominated by a cool blackbody-like component with kT $\lesssim$ 0.1 keV, with a bolometric luminosity $L_{\rm BOL}>10^{39} \rm erg/s$, and little detected emission above 1 keV, the source can also be defined as an ultraluminous supersoft source (ULS or SSUL). The presence of a weak hard tail in the X-ray band in ULSs suggests they are accreting in a super-Eddington regime but seen at \textcolor{black}{high inclinations through} the wind cone obscuring the innermost part of the disc (\citealt{Middleton_2015a}, \citealt{Gurpide_2021b}). \textcolor{black}{This is confirmed by the presence of sharp drops in flux or dips in light-curves followed by spectral softening} (e.g. \citealt{Stobbart2004, Urquhart2016, Alston_2021}). These winds were suggested in several ULXs due to the presence of unresolved and intense features in CCD spectra, typically at lower energies ($<$ 2 keV; \citealt{Stobbart_2006, Sutton_2015, Middleton2014, Middleton2015b}, though recently detected in two cases above 6 keV: \citealt{Walton_2016, Brightman_2023}). Critically, they have since been unambiguously resolved in several ULXs with the XMM-\textit{Newton} Reflection Grating Spectrometers (RGS, hereafter, e.g. \citealt{Pinto_2016,Pinto_2017,Kosec_2018a,Kosec_2018b}). The Doppler blueshift of the absorption lines unveiled the long-sought relativistic ($0.1-0.2\,c$) winds predicted by theoretical simulations of super-Eddington accretion discs (\citealt{Takeuchi2013}); for a recent review on ULX winds see \citet{Pinto_2023}. 


\subsection{HOLMBERG II X-1}
\begin{figure}
	\centering
        \includegraphics[width=0.50\textwidth]{ 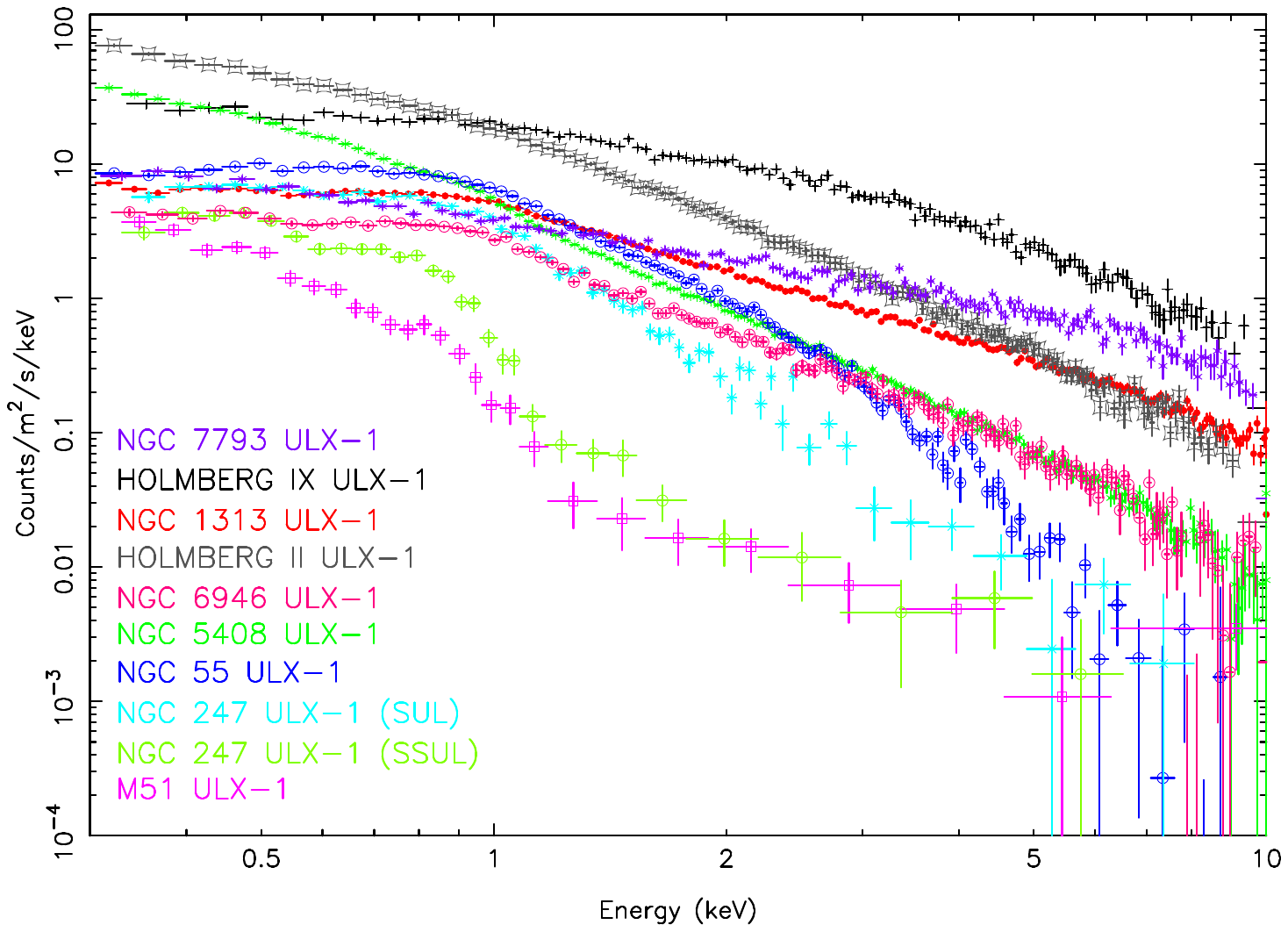}
        \vspace{-0.75cm}
		\caption{{\small Comparison of some ULX XMM-\textit{Newton}/EPIC-PN spectra from the harder to softer (top to bottom). Ho II X-1 spectrum is the brightest below 1 keV. Adapted from \citet{Barra_2022}.}}
		\label{fig: comparison_spectra}
        \vspace{-0.3cm}
\end{figure}

Holmberg II X-1 (hereafter Ho II X-1), located at a distance of 3.05 Mpc\footnote{https://ned.ipac.caltech.edu} in the Holmberg II dwarf galaxy, is characterized by a luminosity in the X-ray band (0.3-10 keV) $L_{\rm X} \ge 10^{39} \rm erg/s$ and up to $10^{40} \rm erg/s$ (\citealt{Gurpide_2021b}). It is possible to see from Fig. \ref{fig: comparison_spectra}
 that the X-ray spectrum of Ho II X-1 tends to fit in between the spectra of bright SUL and the hardest ULXs. 
 
 
Importantly, observations of  Ho II X-1 in the radio reveals collimated jets, which are at least partially responsible for inflating the surrounding nebula, carrying a kinetic luminosity of $>$ 10$^{39}$ erg/s (\citealt{Cseh_2014}). In this sense, Ho II X-1 is expected to be analogous to the extreme Galactic source, SS433 (\citealt{Fabrika2004}), the latter considered to be an edge-on ULX (\citealt{Begelman2006, Middleton2021_SS433}) and proposed to harbour a fairly massive black hole of $\ge$ 25 M$_{\odot}$ (\citealt{Cseh_2014}), in agreement with previous X-ray spectral analysis (\citealt{Goad_2006}). \textcolor{black}{In fact, by comparing the spectral modeling of the microquasar GRS 1915+105 at very high state (e.g, the $\chi$ class) with that of Ho II X-1, a compact object of few tens of Solar masses radiating at or above its Eddington limit was forecasted for the latter case and, however, no more massive of 100 M$_{\odot}$ (\citealt{Goad_2006}).}

\textcolor{black}{A relative uncrowded field around Ho II X-1 permitted the identification of an optical counterpart. In fact, the presence of a He {\scriptsize{II}} $\lambda$4686 nebula, with a luminosity L $ \sim 10^{36} \ \rm erg/s$, and a point-like young companion star consistent with a spectral type O4V or B3 Ib were reported (\citealt{Pakull_2002}, \citealt{Kaaret_2004b}, \citealt{Kaaret_2005}; although see \citealt{Tao_2012}}).
 
The unknown demographic of ULXs, and difficulty to separate them spectrally (see \citealt{Walton_2018, Mills2023}, although several alternatives have been suggested, e.g. \citealt{Middleton2019}, \citealt{Song_2020}, \citealt{Middleton_23}), makes Ho II X-1 an important target given its probable black hole nature. Specifically, we can learn about the nature of the accretion flow by comparing the properties of the wind and evolution of the spectrum to other ULXs and sub-Eddington sources. 
 
This paper is structured as follows; in Sect. \ref{Observations and Spectral modelling} we report on the observations of the source and the spectral modelling in Sect. \ref{Spectral modelling}. In Sect. \ref{High-resolution X-ray spectroscopy} we present results on the lines seen by RGS. In Sect. \ref{Discussion} we discuss our results and give our conclusions in Sect. \ref{Conclusions}. All uncertainties are at 1$\sigma$ (68 \% level).

\section{Observations and data reduction}
\label{Observations and Spectral modelling}
Ho II X-1 was observed by XMM-\textit{Newton} 20 times over a period of 19 years (see Table \ref{table:observations log}), benefiting from a recent 250 ks campain (PI: Middleton). This high quality data-set spread over time allows us to probe both short-term (hours-days) and long-term (months-years) variability. The raw data were obtained from the XMM-\textit{Newton} Science Archive (XSA)\footnote{https://www.cosmos.esa.int/web/XMM-\textit{Newton}/xsa} and were reduced with the \textit{Science Analysis System} ({\scriptsize{SAS}}) version 18.0.0\footnote{https://www.cosmos.esa.int/web/XMM-\textit{Newton}}. \textcolor{black}{In the observations with ID 0112522201 and 0112522301 the satellite's filters were closed and the observation ID 0843840201 is in timing mode and badly calibrated, which leaves 17 well-exposed observations.} We used recent calibration files (October 2022).

\begin{figure*}
		\centering
		\includegraphics[width=0.47\textwidth]{ 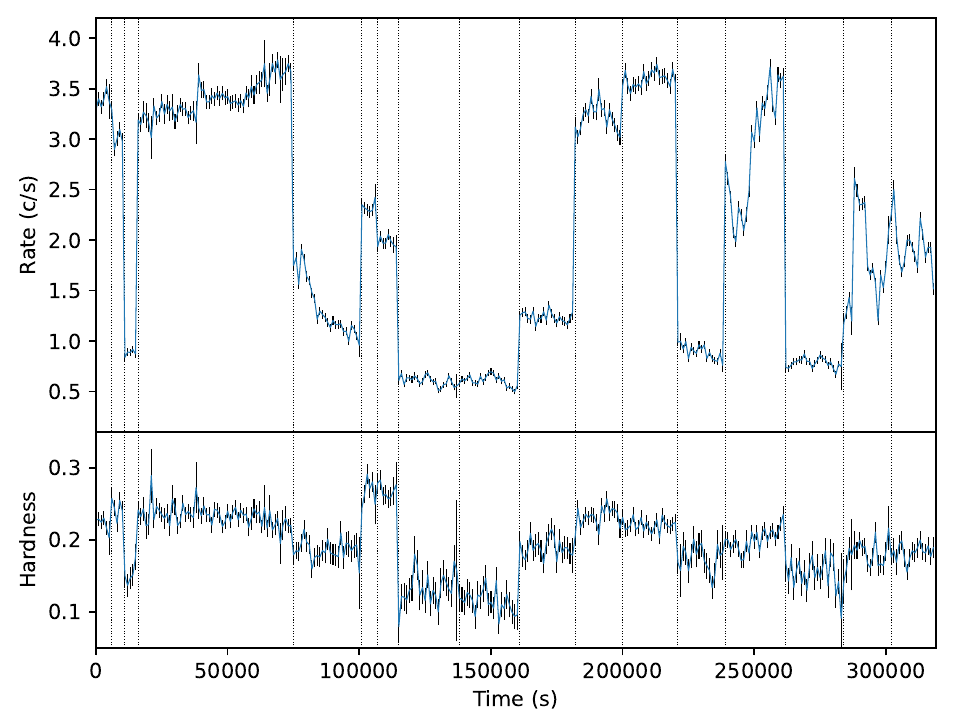}
          \includegraphics[width=0.52\textwidth]{ 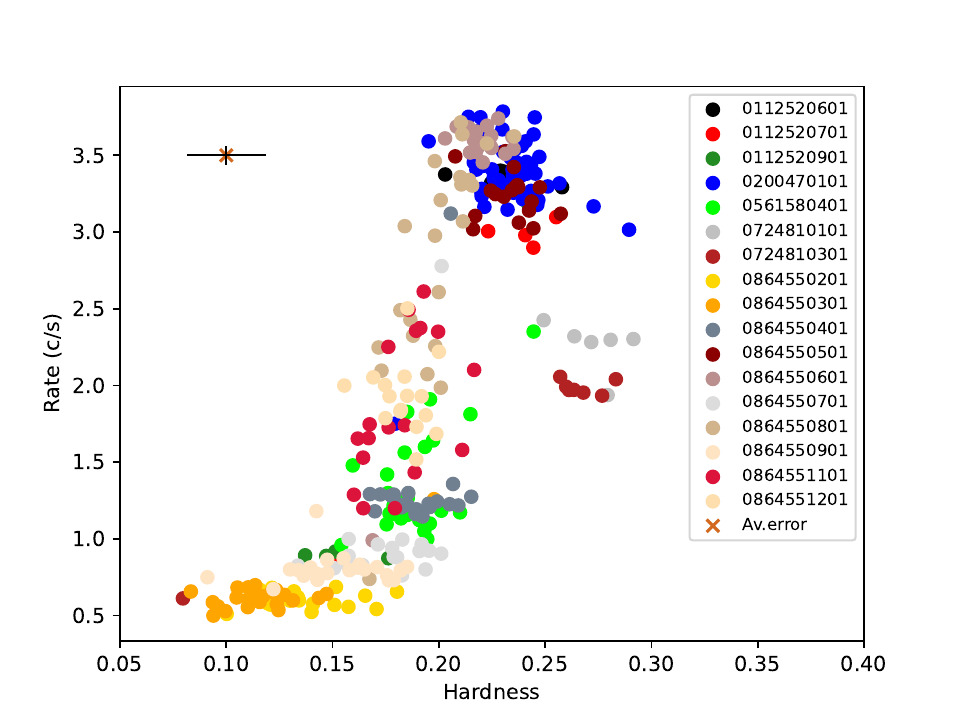}
		\caption{{\small Left panel: Concatenated XMM-\textit{Newton} EPIC-PN lightcurves for the \textcolor{black}{17 well-exposed} observations of Ho II X-1 and the hardness ratio below (here defined as the (1.5-10 keV/ 0.3-10 keV counts ratio) from 2002 to 2021 with time bins of 1 ks. The vertical dotted lines were included in order to separate each of the individual XMM observations. Right panel: hardness-intensity diagram where the points belonging to the same observation have the same colour. 
  }}
		\label{fig: lightcurves}
	\end{figure*}

\begin{figure*}
		\centering
		\includegraphics[width=0.80\textwidth]{ 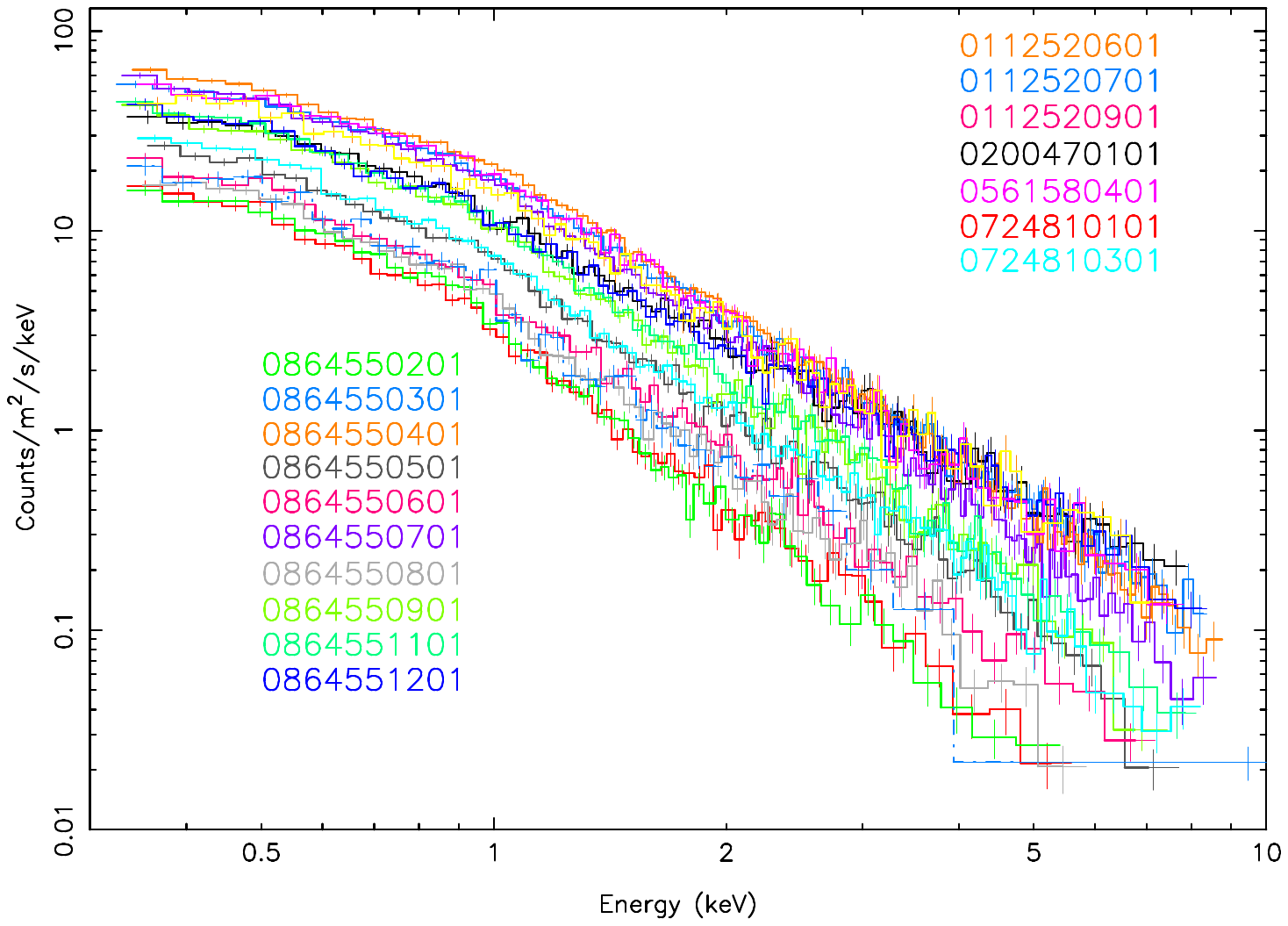}
        \vspace{-0.5cm}
		\caption{{\small EPIC-PN spectra for the 17 \textcolor{black}{well-exposed} observations of Ho II X-1. MOS 1,2 spectra are not shown for clarity purposes.}}
		\label{fig: EPIC SPECTRA}
	\end{figure*}

\begin{center}
	\begin{table}
\caption{Table of the XMM-\textit{Newton} observations of Holmberg II X-1.}  
 \renewcommand{\arraystretch}{1.}
 \small\addtolength{\tabcolsep}{0pt}
 \vspace{0.1cm}
	\centering
	\scalebox{0.9}{%
	\begin{tabular}{ccccccc}
    \toprule
    {{Obs. ID}}  &
    {{Date}}  &
    {{t$\,_{\rm tot}$ [s]}} &
    {{t$\,_{\rm net, \, EPIC-PN}$ [s]}} &
    {{{{CR}$\,_{\rm EPIC-PN}$ [c/s]}}} \\
    
    \midrule
0112520601 & 2002-04-10  & 12640  & 4637  & 2.62 $\pm$ 0.02  \\\midrule
0112520701 & 2002-04-16  & 13869  & 3503  & 2.26 $\pm$ 0.02 \\\midrule
0112520901 & 2002-09-18  & 6888 & 4235  & 0.679 $\pm$ 0.01  \\\midrule
0200470101 & 2004-04-15  & 104677 & 37539  & 2.62 $\pm$ 0.008 \\\midrule
0561580401 & 2010-03-26  & 53852 & 21712  & 1.06 $\pm$ 0.007 \\\midrule
0724810101 & 2013-09-09  & 15268  &  4650  & 1.78 $\pm$ 0.02  \\\midrule
0724810301 & 2013-09-17  & 15000  &  5988  & 1.73 $\pm$ 0.02  \\\midrule
0864550201 & 2021-03-08  & 26300  &  17414  & 0.468 $\pm$ 0.005  \\\midrule
0864550301 & 2021-03-12  & 25900  &  19577  & 0.481 $\pm$ 0.005  \\\midrule
0864550401 & 2021-03-15  & 27000 & 17600 & 0.956 $\pm$ 0.007  \\\midrule
0864550501 & 2021-03-19  & 23000 & 13513  & 2.54 $\pm$ 0.01  \\\midrule
0864550601 & 2021-03-23  & 26000 & 17768  & 2.78 $\pm$ 0.01  \\\midrule
0864550701 & 2021-03-27  & 20500 & 14054  & 0.690 $\pm$ 0.007  \\\midrule
0864550801 & 2021-04-04  & 26000 & 19202  & 2.25 $\pm$ 0.01  \\\midrule
0864550901 & 2021-04-08  & 26000 & 17745  & 0.608 $\pm$ 0.006  \\\midrule
0864551101 & 2021-03-31  & 25300 & 11917  & 1.40 $\pm$ 0.01  \\\midrule
0864551201 & 2021-04-12  & 25000 & 12908  & 1.50 $\pm$ 0.01  \\\midrule
    \bottomrule
    \end{tabular}}\label{table:observations log}
       \vspace{0.3cm}
      \begin{tablenotes}
      \small
  \item[]  t$\,_{\rm net}$ is the exposure time after the removal of periods with high background and {CR}$\,_{\rm EPIC-PN}$ is the net source count rate.
    \end{tablenotes}
\end{table}
\end{center}
The events files of the EPIC camera were generated through the \textit{epproc} and \textit{emproc} tasks and thereafter filtered for the flaring particle background at the recommended threshold above 10 keV (count rate $<$ 0.5 ct/s for EPIC-PN and $<$ 0.35 ct/s for EPIC-MOS 1 and 2). We selected the counts in the source and  background regions using the \textit{evselect} task and corrected the lightcurves with the \textit{epiclccor} task. We chose a circular region for the source of 30 arcsec radius centered on the Chandra X-ray position (RA: 08h 19m 28.99s, Dec: +70d 42m 19.37s) whilst for the background we selected a larger circular region a few arc minutes away from the source, but still on the same chip, avoiding contamination from the copper ring and away from chip gaps. The \textit{rmfgen} and \textit{arfgen} tasks were used to generate response matrices and effective area files. For illustration, the concatenated 0.3-10\,keV EPIC-PN lightcurves of the individual observations are shown in Fig. \ref{fig: lightcurves}. The XMM-\textit{Newton} lightcurves confirm that the source brightness changed by a factor of 2-3 within observations and up to a factor of 6-7 between observations.
Defining the hardness ratio (HR) as the ratio between the counts in the 1.5-10\,keV and the 0.3-10\,keV energy bands, we see that the HR is generally higher (i.e. the spectrum is harder) when the source is brighter and lower in proximity of the dips (Fig. \ref{fig: lightcurves} lower-left and right panels) which is typical in ULXs (e.g. \citealt{Weng2018}). The EPIC-PN spectra of all the observations are shown in Fig. \ref{fig: EPIC SPECTRA} divided by the instrument response but without any spectral model.



The archival RGS spectra were already thoroughly studied in a previous work \citep{Kosec_2018a,Kosec_2021}; we therefore focused on the new data from the 2021 campaign. The RGS data of all observations were reduced according to the standard procedure with the {\textit{rgsproc}} pipeline in {\scriptsize{SAS}}. We filtered out periods of high background by selecting intervals in the lightcurves of the RGS 1,2 CCD\,9 with a count rate below 0.2 ct/s. 

We extracted the $1^{\rm st}$-order RGS spectra in a cross-dispersion region of 0.8 arcmin width, centred on the same source coordinates used for extracting the EPIC spectra. The background regions were chosen by selecting photons outside of 98\% of the source point-spread-function.
ULX RGS spectra typically require at least 100 ks exposure time to achieve the number of counts necessary to detect narrow lines (ideally $\sim$ 10,000 counts, e.g. \citealt{Pinto_2016}).
We stacked the $1^{\rm st}$-order RGS 1 and 2 spectra from the 2021 observations 
with {\textit{rgscombine}} for a total exposure time of 210 ks and about 25,900 net source counts. This is amongst the 10 deepest RGS spectra available for a ULX (for a comparison see Fig. 5 in \citealt{Kosec_2021}). RGS operates in the 0.33-2.5 keV band but the source is above the background (and the instrumental foreground) between 0.4-2.0 keV and we use this band for the analysis. The search for spectral features requires accurate knowledge of the continuum shape, therefore we also stack the 2021 MOS\,1, MOS\,2, and pn spectra to cover the 2-10 keV band. In the end, we produced 4 time-averaged 2021 spectra, one for each camera (RGS, MOS1, MOS2 and pn). 


\section{Broadband X-ray spectroscopy}
\label{Spectral modelling}
\textcolor{black}{For the purpose of studying the behaviour of the source, we initially proceed to analyse the data collected by the CCDs.}
The spectra were modelled with the {\scriptsize{SPEX}} fitting package v3.07.03 (\citealt{kaastra1996_spex}, 2023). EPIC PN and MOS 1,2 spectra were rebinned to at least 1/3 of the spectral resolution and with at least 25 counts per bin using the task {\sc specgroup}. All the models take into account the absorption from the circumstellar and interstellar medium by using the $hot$ model (with a gas temperature of $10^{-4} \ \rm keV$, which yields a quasi-neutral gas phase in {\scriptsize{SPEX}}). All spectral models also account for the source redshift (z=0.00052\footnote{https://ned.ipac.caltech.edu}). EPIC MOS/PN for each observation spectra were fitted simultaneously as they overlap in the 0.3-10 keV band including a multiplicative constant to account for the well known $\lesssim$ 5\% cross-calibration uncertainties. 

	\begin{center}
	\begin{table*}[h!]
\caption{Results from the modelling of the XMM-\textit{Newton} spectra of Holmberg II X-1 with the data of  observation 0200470101. }  
 \renewcommand{\arraystretch}{1.}
 \small\addtolength{\tabcolsep}{-4pt}
 \vspace{0.1cm}
	\centering
	\scalebox{1}{%
	\begin{tabular}{cccccccccccccc}
    \toprule
    {{Parameter /}}  &
    {{RHBB}}  & {{RHBD}} &
    {{RHBM}} & {{RHBCom}} &  {{RHDCom}} &  {{RHMM}} & {{RHDD}} & {{RHBDP}} & {{RHMMGGG}}  \\
    {{component}} & {{Model}}  &  {{Model}}   & {{Model}}  & {{Model}} & {{Model}} & {{Model}}  & {{Model}} & {{Model}} & {{Model}} \\
    
  \midrule
       $L_{X\,bb1}$   &  4.63 $\pm$ 0.07           & 3.53 $\pm$ 0.22             & 3.68 $\pm$ 0.07              & 0.05 $\pm$ 0.02                 &  --- \par                       &  --- \par                  &  --- \par                        & 2.89 $\pm$ 0.06           & --- \par                    \\\midrule
       $L_{X\,bb2}$   &  3.60 $\pm$ 0.12           & --- \par                    & --- \par                     & --- \par                        &  --- \par                       &  --- \par                  &  --- \par                        & --- \par                  & --- \par                     \\\midrule
       $L_{X\,mbb1}$  &  --- \par                  & --- \par                    & 4.8 $\pm$ 0.1                & --- \par                        &  --- \par                       & 5.09 $\pm$ 0.20            &  --- \par                        & --- \par                  & 5.04 $\pm$  0.20            \\\midrule
       $L_{X\,mbb2}$  &  --- \par                  & --- \par                    & --- \par                     & --- \par                         &  --- \par                       & 4.02 $\pm$ 0.17            &  --- \par                       & --- \par                  & 4.02 $\pm$ 0.16             \\\midrule
       $L_{X\,dbb1}$  &  --- \par                  & 4.96 $\pm$ 0.06             &  --- \par                    & --- \par                         & 0.29 $\pm$ 0.18                  & --- \par                  &  4.96 $\pm$ 0.30                & 4.2 $\pm$ 0.3             & --- \par                     \\\midrule
       $L_{X\,dbb2}$  &  --- \par                  & --- \par                    &  --- \par                    & --- \par                         & --- \par                         & --- \par                  & 4.31 $\pm$ 0.26                 & --- \par                  & --- \par                    \\\midrule
       $L_{X\,comt}$  & --- \par                   & --- \par                    & --- \par                     & 8.7 $\pm$ 0.8                    & 9.17 $\pm$ 2.19                  & --- \par                  & --- \par                        & --- \par                  & --- \par                     \\\midrule 
       $L_{X\,gauss}$ & --- \par                   & --- \par                    & --- \par                     & --- \par                        & --- \par                         & --- \par                  & --- \par                         &  --- \par                 & 0.05 $\pm$ 0.08               \\\midrule 
       $L_{X\,pow}$   & --- \par                   & --- \par                    & --- \par                     & --- \par                        & --- \par                         & --- \par                   & --- \par                        &  3.4 $\pm$ 0.1            & --- \par                     \\\midrule
       kT$_{bb1}$     & 0.217 $\pm$ 0.001          &  0.206 $\pm$ 0.001          & 0.204 $\pm$ 0.001            & 0.144 $\pm$ 0.001               & --- \par                        & --- \par                    & --- \par                        & 0.194 $\pm$ 0.005         & --- \par                     \\\midrule 
       kT$_{bb2}$     & 0.783 $\pm$ 0.006          & --- \par                    & --- \par                     & --- \par                        & --- \par                        & --- \par                    & --- \par                        & --- \par                  & --- \par                    \\\midrule
       kT$_{mbb1}$    & --- \par                   & --- \par                    & 1.33 $\pm$ 0.01              & --- \par                        & --- \par                        & 0.369 $\pm$ 0.005           & --- \par                        & --- \par                  & 0.375 $\pm$ 0.006           \\\midrule
       kT$_{mbb2}$    & --- \par                   & --- \par                    & --- \par                     & --- \par                        & --- \par                         & 1.61 $\pm$ 0.02             & --- \par                       & --- \par                  & 1.62 $\pm$ 0.03             \\\midrule
       kT$_{dbb1}$    & --- \par                   & 2.35 $\pm$ 0.02             &  --- \par                    & --- \par                        & 0.135 $\pm$ 0.005               & --- \par                    & 0.63 $\pm$ 0.01                 & 1.62  $\pm$ 0.03          & --- \par                    \\\midrule 
       kT$_{dbb2}$    & --- \par                   & --- \par                    &  --- \par                    & --- \par                         & --- \par                        & --- \par                    & 2.77 $\pm$ 0.04                & --- \par                   & --- \par                      \\\midrule
       kT$_{seed}$    & --- \par                   & --- \par                    & --- \par                     &  0.144 (coupled)                 & 0.135 (coupled)                 & --- \par                    & --- \par                        & --- \par                  & --- \par                      \\\midrule 
       kT$_{e}$       & --- \par                   & --- \par                    & --- \par                     &  2.76 $\pm$ $_{0.28}^{0.35}$     & 3.98 $\pm$ $_{0.72}^{1.27}$      & --- \par                    & --- \par                       & --- \par                   & --- \par                   \\\midrule
       $\tau$         & --- \par                   & --- \par                    & --- \par                     & 4.27 $\pm$ 0.35                  & 3.24 $\pm$ $_{0.79}^{0.40}$      & --- \par                    & --- \par                       & --- \par                   & --- \par                    \\\midrule 
       $\Gamma$       & --- \par                   & --- \par                    & --- \par                     & --- \par                         & --- \par                        & --- \par                    & --- \par                       & 0.59                       & --- \par                       \\\midrule 
       $\tau_{0}$     & --- \par                   & --- \par                    & --- \par                     & --- \par                         & --- \par                        & --- \par                    & --- \par                       & 0.1265                     & --- \par                       \\\midrule 
       $N_{H}$        & 0.500 $\pm$ $_{0}^{0.001}$ & 0.500 $\pm$ $_{0}^{0.001}$ & 0.500 $\pm$ $_{0}^{0.001}$   & 0.50 $\pm$  $_{0}^{0.03}$         & 0.80 $\pm$ 0.01                  & 0.65 $\pm$ 0.03            & 0.70 $\pm$ 0.02                & 0.500 $\pm$ $_{0}^{0.004}$ & 0.65 $\pm$ 0.03                \\\midrule
       $E_{0}^{1}$    & --- \par                   & --- \par                   & --- \par                     & --- \par                          & --- \par                        & --- \par                   & --- \par                       & --- \par                   & 0.529 $\pm$ 0.007               \\\midrule
       $E_{0}^{2}$    & --- \par                   & --- \par                   & --- \par                     & --- \par                          & --- \par                         & --- \par                  & --- \par                       & --- \par                   & 0.80 $\pm$ 0.10                 \\\midrule
       $E_{0}^{3}$    & --- \par                   & --- \par                   & --- \par                     & --- \par                          & --- \par                         & --- \par                  & --- \par                       & --- \par                   & 0.90 $\pm$ 0.01                  \\\midrule
        FWHM          & --- \par                   & --- \par                   & --- \par                     & --- \par                          & --- \par                         & --- \par                  & --- \par                       & --- \par                   & 0.001                           \\\midrule
   $\rm Norm_{Pow}$   & --- \par                   & --- \par                   & --- \par                     & --- \par                          & --- \par                         & --- \par                  & --- \par                       & 1161 $\pm$ 40          & --- \par                        \\\midrule
   $\rm Norm_{ 1}$    & --- \par                   & --- \par                   & --- \par                     & --- \par                         & --- \par                        & --- \par                  & --- \par                        & --- \par                   & 622 $\pm$  $_{225.49}^{1609}$     \\\midrule
   $\rm Norm_{ 2}$    & --- \par                   & --- \par                   & --- \par                     & --- \par                        & --- \par                         & --- \par                  & --- \par                        & --- \par                   & -34.4 $\pm$  $_{54.2}^{34.4}$     \\\midrule 
   $\rm Norm_{ 3}$    & --- \par                   & --- \par                   & --- \par                     & --- \par                        & --- \par                          & --- \par                  & --- \par                        & --- \par                   & 165 $\pm$ 39                    \\\midrule 
  ${\chi}^2$/ d.o.f   & 1978/404                   & 1056/404                   & 1102/404                     & 558/403                         & 545/403                           & 640/404                   & 661/404                         & 622/403                    & 597/398                         \\\midrule
    ${\chi}^2_{PN}$   & 1001                       & 478                        & 504                           & 187                            & 166                               & 227                       & 256                               & 222                      & 209                             \\\midrule
  ${\chi}^2_{MOS1}$   & 472                        & 267                        & 279                           & 163                            & 167                               & 179                       & 185                               & 180                      & 166                             \\\midrule
  ${\chi}^2_{MOS2}$   & 505                        & 311                        & 319                           & 208                            & 212                               & 234                       & 240                               & 220                      & 223                             \\\midrule
    \bottomrule       
    \label{table: XMM 0200470101 spectral fits}   
    \end{tabular}}    
    \begin{tablenotes}
      \small
  \item[] Parameter units: $E_{0}^{1}$, $E_{0}^{2}$ and $E_{0}^{3}$ (in keV unit) refer to the centroids of each gaussian line. 
$\rm Norm_{1}$, $\rm Norm_{2}$ and $\rm Norm_{3 } $\ (in $10^{44} \rm ph/s $ units) are their normalisations. The temperatures kT (for each model) and FWHM are expressed in keV unit. The X-ray and bolometric luminosities $L_{\rm X}$ and $L_{\rm BOL}$  (always intrinsic or unabsorbed) are calculated, respectively, between the 0.3 - 10 keV and 0.001 - 1000 keV bands, and are expressed in $10^{39}$ erg/s unit. The column density of the cold gas $N_{H}$ is in 10$^{21}$/cm$^{2}$ unit. $\tau$ is the optical depth of the comptonisation component. $\Gamma$ and $\rm Norm_{Pow}$ are the slope and the normalization (expressed in  $10^{44} \ \rm ph/s \ (keV $ units) of the powerlaw component. $\tau_{0}$ is the optical depth of the cutoff component at E=7.9 keV (for more information see the {\scriptsize{SPEX}} manual).
    \end{tablenotes}
 \vspace{-0.3cm}
\end{table*}
\end{center}

\subsection{Testing different spectral models for Obs. 02004710101}
In order to describe Ho II X-1 spectrum, we proceed to test several models on the longest observation (Obs. ID 0200470101), which provides the largest number of counts,  with double thermal models, as common procedure in ULX spectra modelling (\citealt{Gurpide_2021a}, \citealt{Pintore_2015},  \citealt{Stobbart_2006}, \citealt{Walton_2018}). In particular we want to compare our results with those obtained by \cite{Walton_2015} and \cite{Gurpide_2021a}.

The spectral models that we used to describe the data were different combinations of the following  components: blackbody emission (\textit{bb}) and blackbody modified by coherent Compton scattering (\textit{mbb}) \textcolor{black}{often used for super-Eddington discs (\citealt{Pinto_2017,Barra_2022})}, multi-temperature disc blackbody (\textit{dbb}) and inverse comptonisation of soft photons in a hot plasma (\textit{comt}), \textcolor{black}{typically used for Eddington-limited Galactic XRBs but sometimes also for ULXs (\citealt{Done2007,Middleton_2015a})}. In order to fit emission and absorption lines we used gaussian lines. A summary of the models tested is below. All models are corrected for the redshift and neutral interstellar absorption.
\begin{itemize}
  \item RHB(B): One or more blackbody emission. 
  \vspace{0.1cm}
  \item RHBD: cool blackbody emission and a warmer disc blackbody.
  
  \vspace{0.1cm}
   \item RHDD: a double disc- blackbody model.
   \vspace{0.1cm}
  \item RHBM: a modified single blackbody model.
   \vspace{0.1cm}
   \item RHMM: a modified double blackbody model.
   \vspace{0.1cm}
  \item RHBCom: blackbody plus comptonisation. 
   \vspace{0.1cm}
   \item RHDCom: disc blackbody plus comptonisation. 
   \vspace{0.1cm}
  \item RHBDP: a powerlaw with a cutoff (\textit{etau}) is added to the RHBD model. 
   \vspace{0.1cm}
  \item RHMMG(G): One or more gaussian lines are included to the blackbody fit to account for emission or absorption lines. 
\end{itemize}

Fits to the data with continuum-only models (i.e. without gaussians) show residuals around 1 keV, 0.75 keV and 0.5 keV (as previously reported in several ULXs including Ho II X-1: \citealt{Middleton2015b}). These lines are resolved with the high-resolution RGS spectrometers, as explained below. 

\begin{figure}
		\centering
		\includegraphics[width=0.50\textwidth]{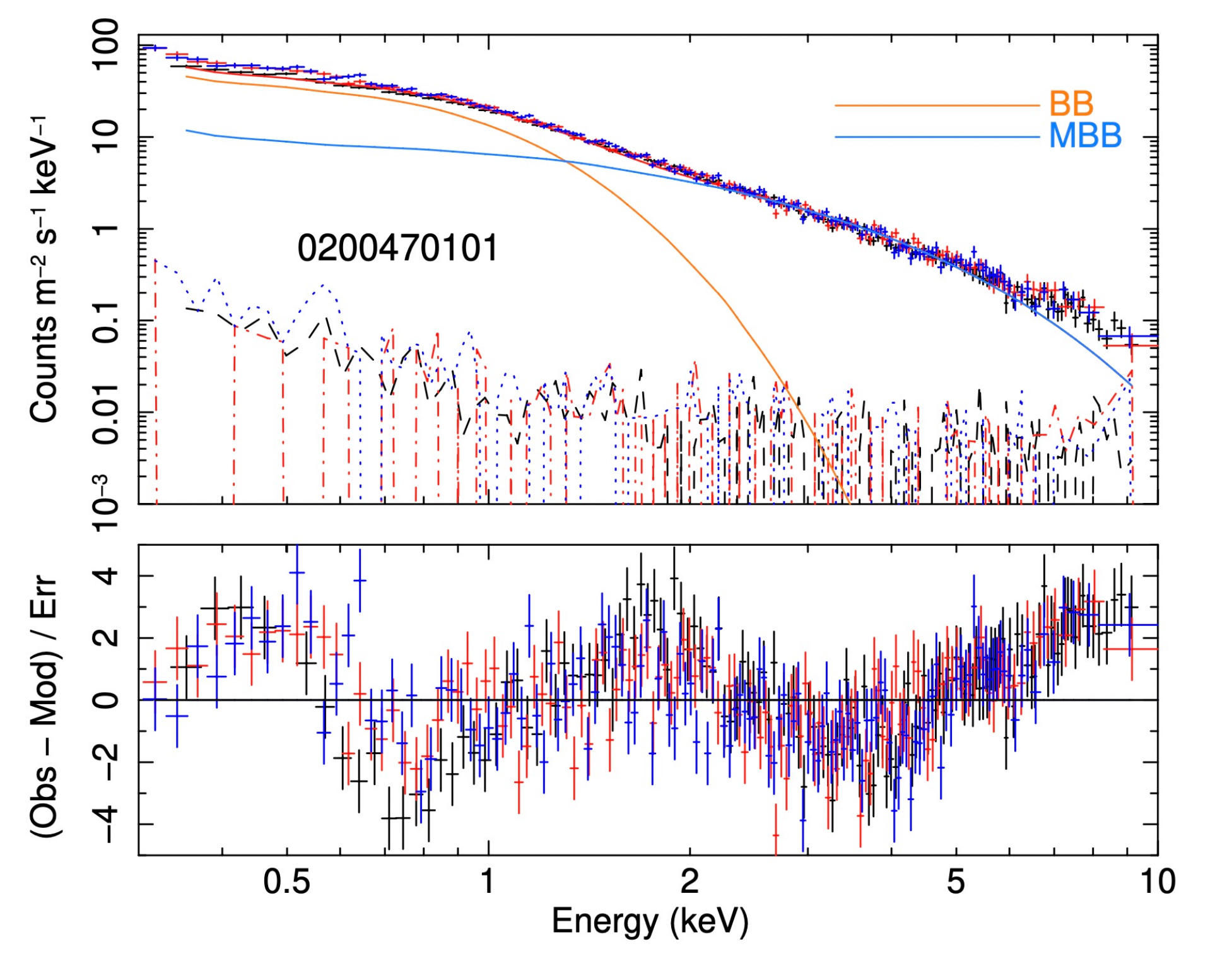}
  \includegraphics[width=0.53\textwidth]{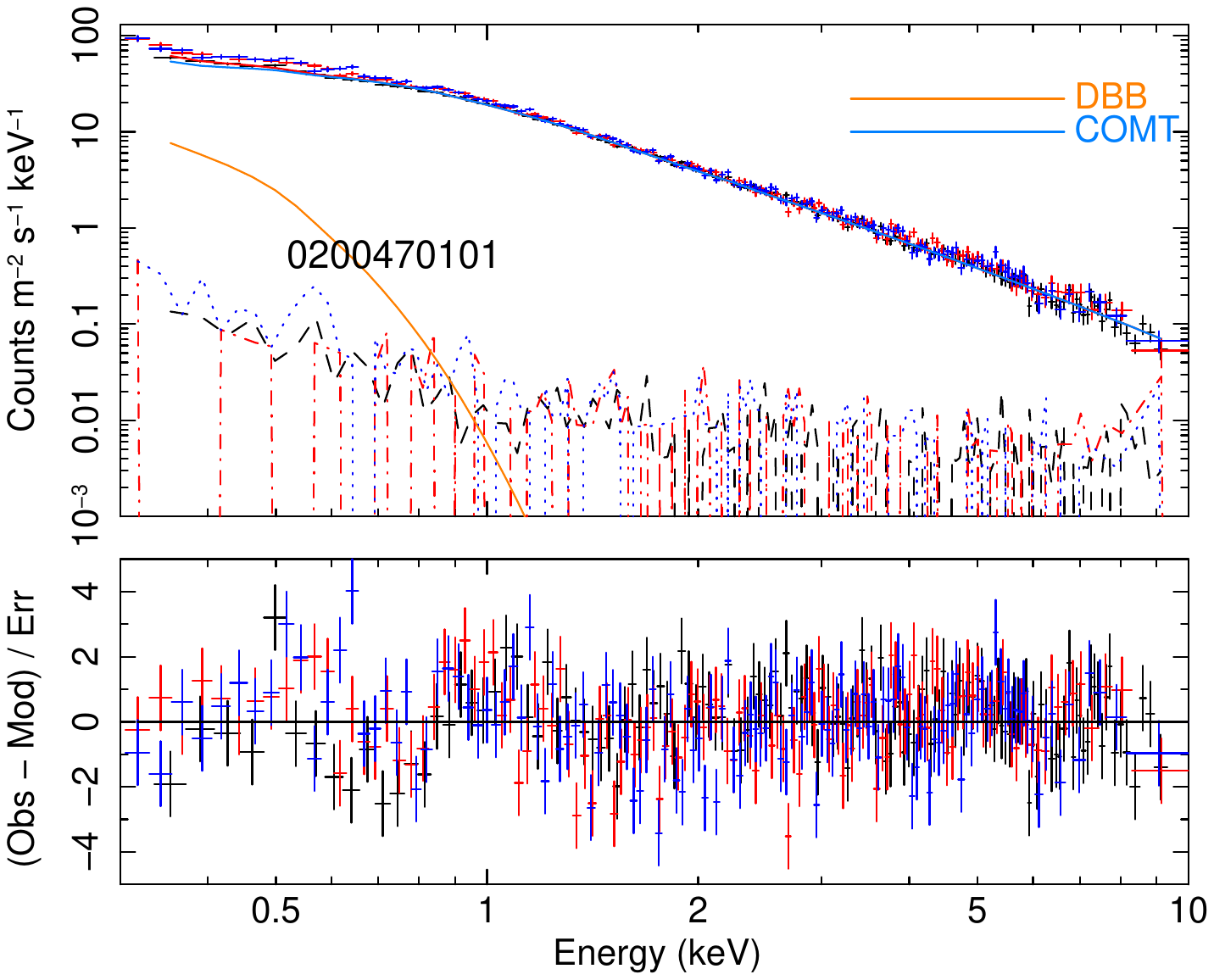}
  \includegraphics[width=0.50\textwidth]{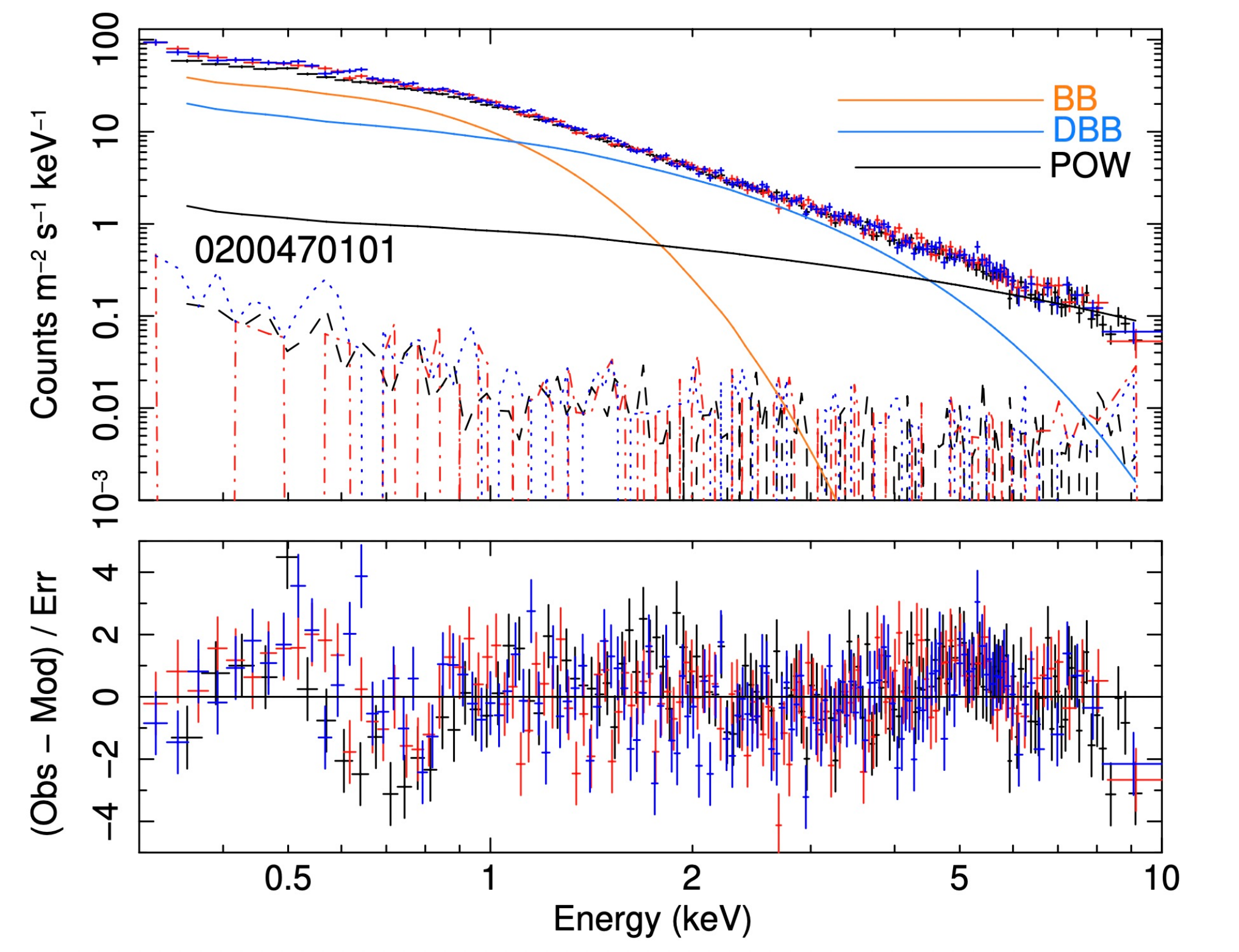}
        \vspace{-0.3cm}
		\caption{{\small Spectral fits to the EPIC PN and MOS 1,2 data using the RHBM model (top panel), the RHDCom (middle panel) and the RHBDP models (bottom panel) for the Obs. ID 0200470101. The dashed dotted lines indicate the background spectra. }}
		\label{fig: best fits model for the obs. 0200470101}
	\end{figure}
\begin{figure}
		\centering
  \includegraphics[width=0.55\textwidth]{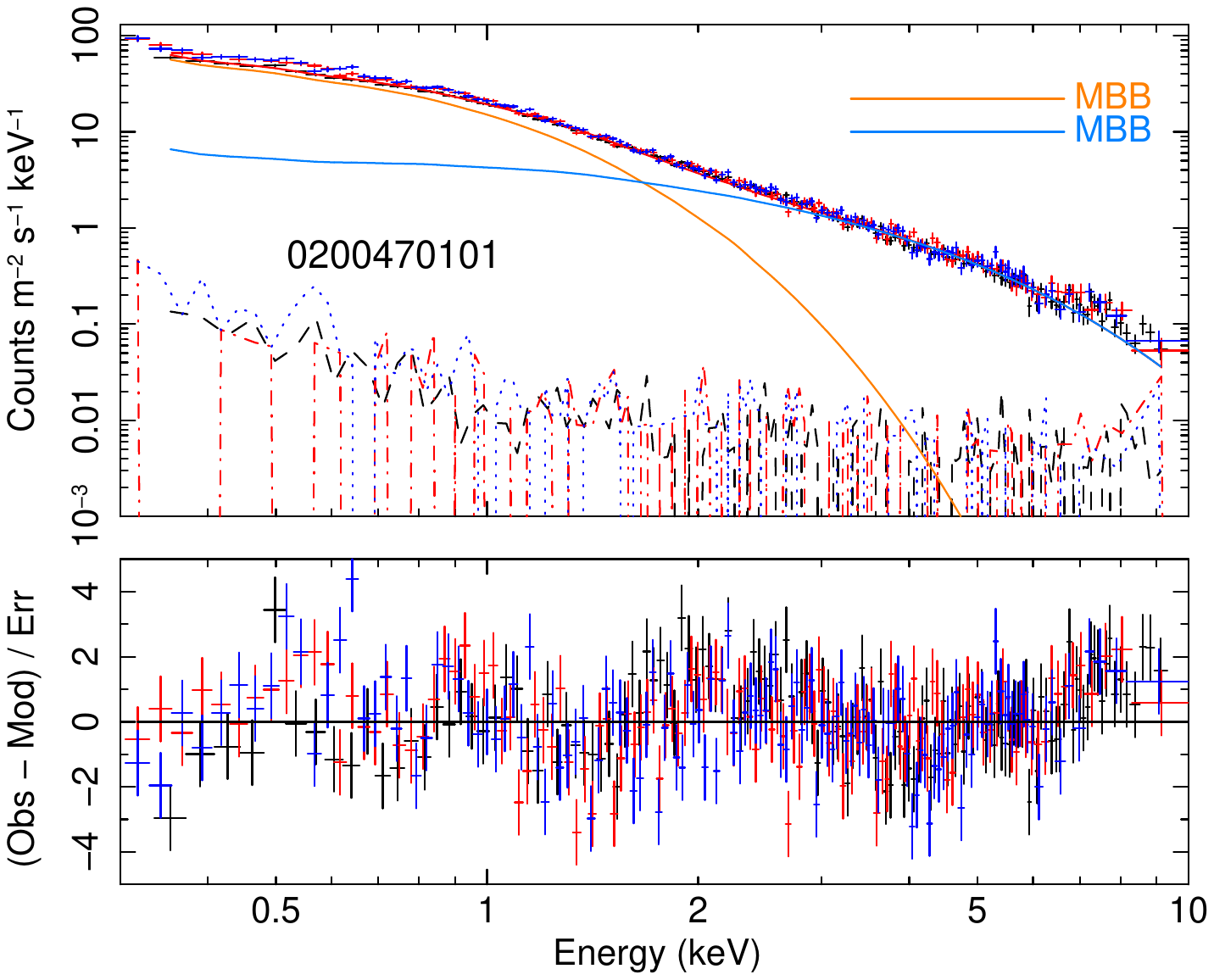}
  \includegraphics[width=0.55\textwidth]{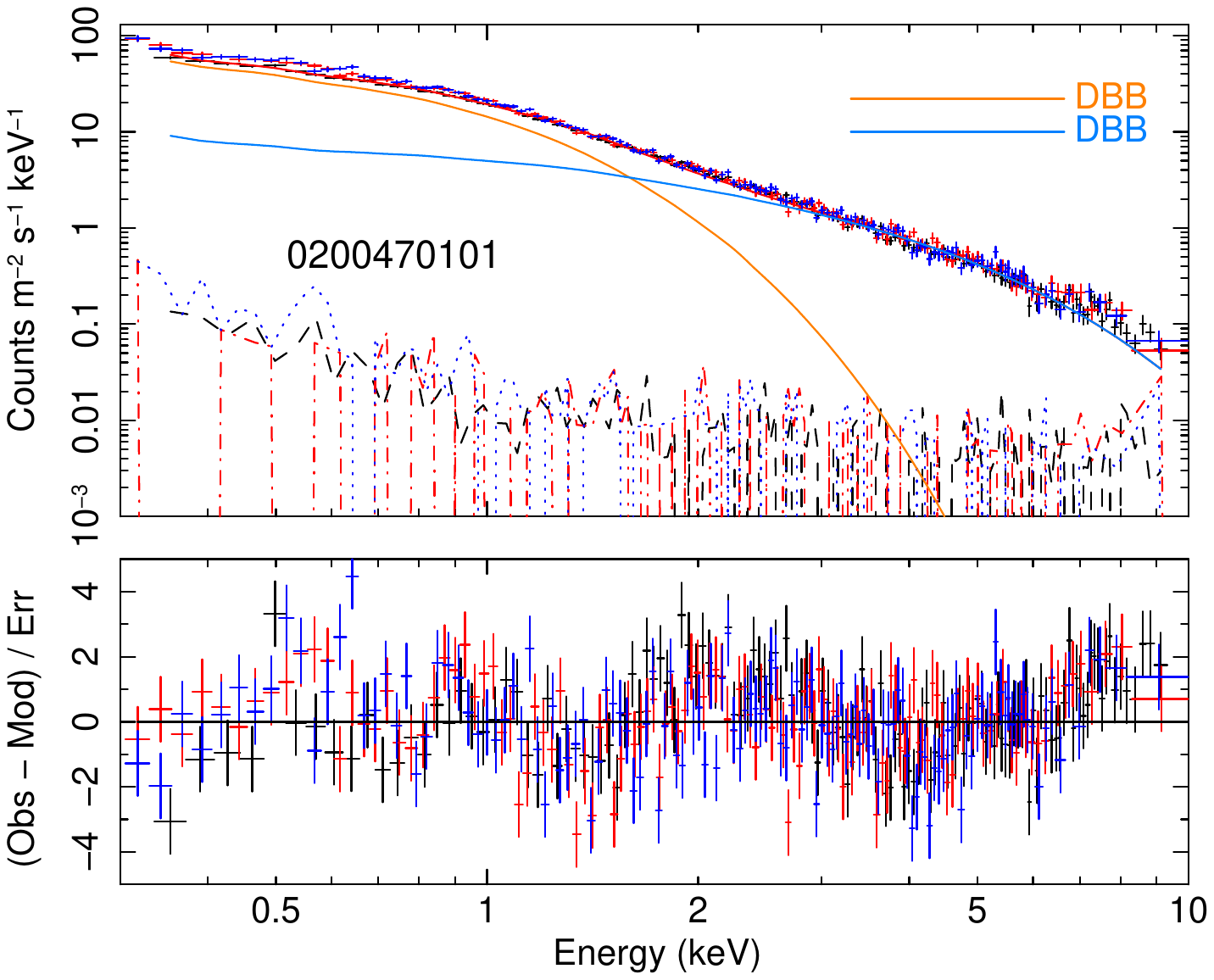}
        \vspace{-0.3cm}
		\caption{{\small Spectral fits to the EPIC PN and MOS 1,2 data using the RHMM model (top panel) and RHDD model (bottom panel) for the Obs. ID 0200470101. }}
		\label{fig: best fits model for the Obs. 0200470101 or RHMM and RHDD models}
	\end{figure}

The spectral fits of Obs. ID 0200470101 are reported in Fig. \ref{fig: best fits model for the obs. 0200470101} and \ref{fig: best fits model for the Obs. 0200470101 or RHMM and RHDD models}. In particular, the double thermal component \textit{bb+mbb} (RHBM model, top panel) does not give a good description of the data above 6 keV. For the power law in the \textit{bb+dbb+pow} model (RHBDP) we adopted a cutoff (\textit{etau}) at 7.9 keV and a spectral index of 0.59 (\citealt{Walton_2020}).
Although the RHDCom and RHBDP models  provide a slightly better fit, we prefer to continue with the \textit{mbb+mbb} model (RHMM model, Fig. \ref{fig: best fits model for the Obs. 0200470101 or RHMM and RHDD models}, top panel), with temperatures of $\sim$ 0.37 keV and $\sim$ 1.6 keV for the cool and hot component respectively, in order to study the behaviour of the thermal components with the bolometric luminosity (Fig. \ref{fig: L-T plot}) and the radii. The latter model provides also a lower ${\chi}^2$ with respect to the \textit{dbb+dbb} model (RHDD model, Fig. \ref{fig: best fits model for the Obs. 0200470101 or RHMM and RHDD models}, bottom panel). \textcolor{black}{The cool (hot) thermal component accounts for the mainly soft (hard) X-ray emission, coming from the wind and the outer (inner) part of the disc.}



The results from our fits of Obs. ID 02004710101 are shown in Table \ref{table: XMM 0200470101 spectral fits}. \textcolor{black}{We also test the other models mentioned in the Table \ref{table: XMM 0200470101 spectral fits} but we were unable to constrain the parameters of the hard tail of the comptonisation (see Sect. \ref{Spectral modelling of individual observations}). This is likely due to the lack of high-energy coverage (E $>$10 keV) during this epoch. By adding, in fact, NuSTAR data, an additional component (\textit{comt} or \textit{pow}) is needed to describe the spectrum at high energies (see Sect.  \ref{X-ray broadband spectroscopy}).} The total column density in several models assumes values, particularly for the RHBDP model (see \ref{table: Table results for the best fit RHBDP model for all observations. }, perhaps due to the emission lines at 0.5 keV), smaller than the Galactic value (see Table \ref{table: XMM 0200470101 spectral fits}). So we decided to restrict the column density to have a minimum value of $5 \times 10^{20} \ \rm cm^{-2}$.

\subsection{Spectral modelling of individual observations}

\label{Spectral modelling of individual observations}
We show in Fig. \ref{fig: EPIC SPECTRA} the EPIC PN spectra of the 17 observations analysed. As with the longest observation, we fitted the EPIC PN and MOS 1,2 spectra of all observations with all models described before. The primary goal is to understand the trends of the spectral components with the time and, then, to check for systematic effects on the results from the model choice.
\textcolor{black}{In particular, testing several models, we obtained $N_{\rm{H}}$ values ranging from $(0.5-0.8) \times10^{21} \rm cm^{-2}$. This was likely due to some dependence with the model applied (i.e. the curvature adopted in the soft band). For clarity and comparison with other sources, we focus on the best-fit RHMM model in the main text. The first fits with $N_{\rm{H}}$ free yielded values consistent within 2-3\,$\sigma$. We therefore rerun them by fixing the column density at the average value of $7\times10^{20} \rm cm^{-2}$}. 
The results are shown in Table \ref{table: Table results for the best fit RHMM model for all observations.  } for all the observations. 
The best-fit RHMM model provided a good representation of the data with the exception of the well known residuals at 0.5 keV, 0.75 keV and 1 keV. However, we stress that these residuals do not dramatically vary using the models with comptonisation (see e.g Fig. \ref{fig: best fits model for the obs. 0200470101}) and are later fitted by adding three gaussian lines to the RHMM model (see section \ref{sec:wind_variability}).

\begin{center}
	\begin{table*}
\caption{Results of the best fit RHMM model for all observations.}  
 \renewcommand{\arraystretch}{1.}
 \small\addtolength{\tabcolsep}{1pt}
 \vspace{0.1cm}
	\centering
	\scalebox{.95}{%
	\begin{tabular}{cccccccccccc}
    \toprule
    {{Obs. ID}}  &
    {{$\rm{N_{H}}$}}  &
    {$\rm{kT_{mbb1}}$} [keV] &
    {$\rm{kT_{mbb2}}$} [keV] &
    {$\rm{L_{X\,mbb1}}$} [erg/s] & 
    {$\rm{L_{X\,mbb2}}$} [erg/s] & 
    ${\chi}^2$/ d.o.f &
    ${\chi}^2_{\rm{PN}}$ &
    ${\chi}^2_{\rm{MOS1}}$ &
    ${\chi}^2_{\rm{MOS2}}$ \\
    
    \midrule
0112520601 & 0.7 & 0.373 $\pm$ 0.007 & 1.64   $\pm$ 0.07 & 5.18 $\pm$ 0.19 & 3.70 $\pm$ 0.23 &  257/258 & 95  & 68  & 94  &   \\\midrule
0112520701 & 0.7 & 0.336 $\pm$ 0.007 & 1.78   $\pm$ 0.07 & 4.23 $\pm$ 0.20 & 3.99 $\pm$ 0.40 &  221/218 & 100 & 74  & 47  &   \\\midrule
0112520901 & 0.7 & 0.248 $\pm$ 0.008 & 1.05   $\pm$ 0.06 & 1.38 $\pm$ 0.12 & 0.79 $\pm$ 0.13 &  119/119 & 57  & 25  & 37  &   \\\midrule
0200470101 & 0.7 & 0.361 $\pm$ 0.002 & 1.58   $\pm$ 0.02 & 5.19 $\pm$ 0.07 & 4.08 $\pm$ 0.14 &  642/405 & 233 & 179 & 230 &   \\\midrule
0561580401 & 0.7 & 0.278 $\pm$ 0.003 & 1.34   $\pm$ 0.03 & 2.09 $\pm$ 0.06 & 1.47 $\pm$ 0.07 &  326/293 & 138 & 69  & 119 &   \\\midrule
0724810101 & 0.7 & 0.322 $\pm$ 0.007 & 2.04   $\pm$ 0.08 & 3.23 $\pm$ 0.15 & 3.98 $\pm$ 0.31 &  243/255 & 76  & 84  & 83  &   \\\midrule
0724810301 & 0.7 & 0.305 $\pm$ 0.006 & 1.94   $\pm$ 0.06 & 3.04 $\pm$ 0.15 & 3.96 $\pm$ 0.28 &  256/256 & 85  & 93  & 78  &   \\\midrule
0864550201 & 0.7 & 0.239 $\pm$ 0.005 & 0.99   $\pm$ 0.04 & 1.06 $\pm$ 0.05 & 0.49 $\pm$ 0.05 &  190/169 & 66  & 67  & 57  &   \\\midrule
0864550301 & 0.7 & 0.254 $\pm$ 0.005 & 1.02   $\pm$ 0.05 & 1.17 $\pm$ 0.05 & 0.39 $\pm$ 0.06 &  235/160 & 89  & 51  &  95 &   \\\midrule
0864550401 & 0.7 & 0.286 $\pm$ 0.005 & 1.34   $\pm$ 0.04 & 1.94 $\pm$ 0.07 & 1.37 $\pm$ 0.10 &  279/226 & 113 & 85  & 81  &   \\\midrule
0864550501 & 0.7 & 0.354 $\pm$ 0.004 & 1.69   $\pm$ 0.04 & 4.90 $\pm$ 0.11 & 4.16 $\pm$ 0.21 &  357/328 & 112 & 125 & 120 &   \\\midrule
0864550601 & 0.7 & 0.351 $\pm$ 0.004 & 1.53   $\pm$ 0.03 & 5.50 $\pm$ 0.11 & 4.23 $\pm$ 0.21 &  441/344 & 159 & 162 & 120 &   \\\midrule
0864550701 & 0.7 & 0.256 $\pm$ 0.006 & 1.19   $\pm$ 0.04 & 1.39 $\pm$ 0.08 & 0.99 $\pm$ 0.09 &  225/192 & 101 & 70  & 54  &   \\\midrule
0864550801 & 0.7 & 0.315 $\pm$ 0.003 & 1.34   $\pm$ 0.02 & 4.45 $\pm$ 0.10 & 3.26 $\pm$ 0.16 &  457/320 & 190 & 123 & 144 &   \\\midrule
0864550901 & 0.7 & 0.264 $\pm$ 0.005 & 1.12   $\pm$ 0.04 & 1.32 $\pm$ 0.06 & 0.70 $\pm$ 0.07 &  228/193 & 112 & 54 & 62   &   \\\midrule
0864551101 & 0.7 & 0.292 $\pm$ 0.004 & 1.31   $\pm$ 0.03 & 2.89 $\pm$ 0.10 & 1.86 $\pm$ 0.13 &  285/249 & 113 & 84 & 88   &   \\\midrule
0864551201 & 0.7 & 0.287 $\pm$ 0.004 & 1.30   $\pm$ 0.03 & 3.08 $\pm$ 0.10 & 2.04 $\pm$ 0.14 &  313/272 & 108 & 81 & 124  &   \\\midrule
    \bottomrule
    \end{tabular}}
    \label{table: Table results for the best fit RHMM model for all observations. }
       \vspace{0.3cm}
        \begin{tablenotes}
      \small 
  \item[] The units of the parameters are the same of Table \ref{table: XMM 0200470101 spectral fits}. The total column density $N_{\rm H}$ is fixed to the average value  $7\times10^{20} \ \rm cm^{-2}$. 
      \end{tablenotes}
     \vspace{-0.3cm}
\end{table*}
\end{center}

\subsection{Variability of the wind features}
\label{sec:wind_variability}
\begin{figure}
		\centering
		\includegraphics[width=0.50\textwidth]{ 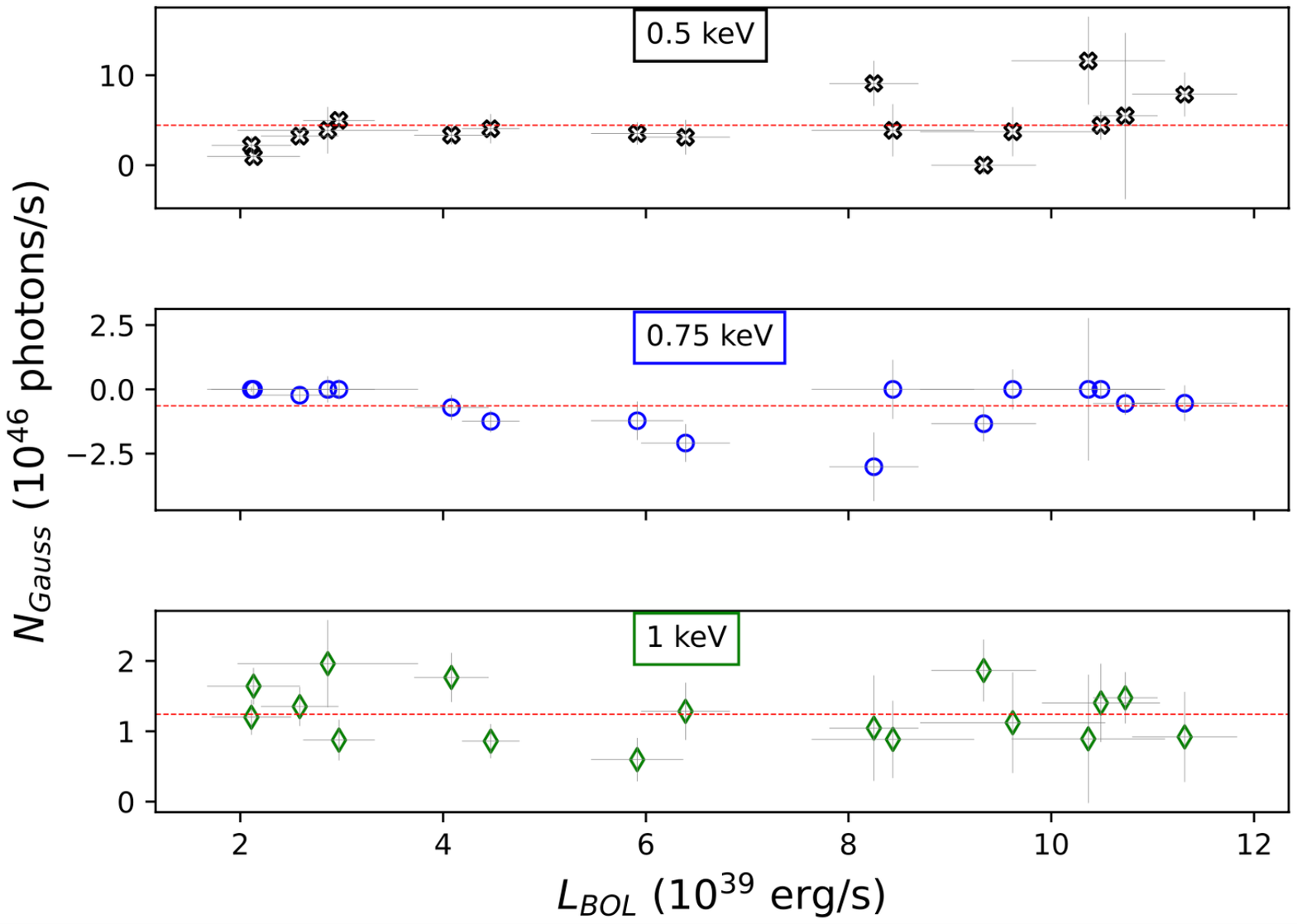}
		\caption{\small{Normalisation of the gaussian lines vs bolometric luminosity for the 3-gaussian model (RHMMGGG) fits. The negative normalisations refer to absorption lines. \textcolor{black}{The red dotted lines in each panel refer to the average value of the gaussian normalization.}}}
		\label{fig: Normalization_vs_Bolometric Luminosity plot}
		\end{figure}

 The strength of the spectral features and their variability in Holmberg II X-1 were studied by using the \textit{mbb} + \textit{mbb} model by adding three gaussian lines (RHMMGGG model). We fixed the line width to 1 eV for all three gaussian lines since EPIC spectra lack the necessary spectral resolution around 1 keV. The energy centroids were left free to vary and they agreed within the error bars with 0.75 for the absorption line, 0.5 and 1.0 keV for the emission lines between the various observations.
The normalisations of the gaussian lines for the seventeen spectra are shown in Fig. \ref{fig: Normalization_vs_Bolometric Luminosity plot}. They do not show significant variability within the uncertainties; the 0.5 keV emission line weakly brightens whilst the 0.75 keV feature is undetected at the highest luminosities.
 
We preferred to use the simplified RHMM(GGG) model in order to avoid complicating the continuum model by adding the powerlaw component for the hard X-ray tail \textcolor{black}{because for such model and the one with the comptonisation several parameters remain unconstrained}. However, we have tested the addition of three gaussian lines to the RHBDP model over the 17 spectra and obtained results consistent with those of the RHMMGGG.

\section{High-resolution X-ray spectroscopy}
\label{High-resolution X-ray spectroscopy}

In order to search for and identify narrow lines, we regrouped the 2021 time-averaged EPIC PN, MOS1, MOS2 and RGS spectra according to the optimal binning discussed in \citet{Kaastra_2016} which adopts energy bins of at least 1/3 of the spectral resolutions and a minimal, positive, number of counts per bin. In this analysis we therefore fit the data by reducing the Cash statistic ($C$-stat; \citealt{Cash_1979}).
Our binning choice smooths the background spectra in the energy range with low statistics,
and removes narrow spurious features. We notice however that both the RGS and EPIC stacked spectra have high statistics and therefore the regrouping turns out to be similar to the standard 1/3 of the spectral resolution in the region of interest for the narrow lines (0.4-2 keV). Moreover, further tests that were performed by rebinning to at least 25 counts per bin, using the {\scriptsize{SAS}} task {\textit{specgroup}} showed consistent results.

\begin{figure}
		\centering
		\includegraphics[width=0.48\textwidth]{ 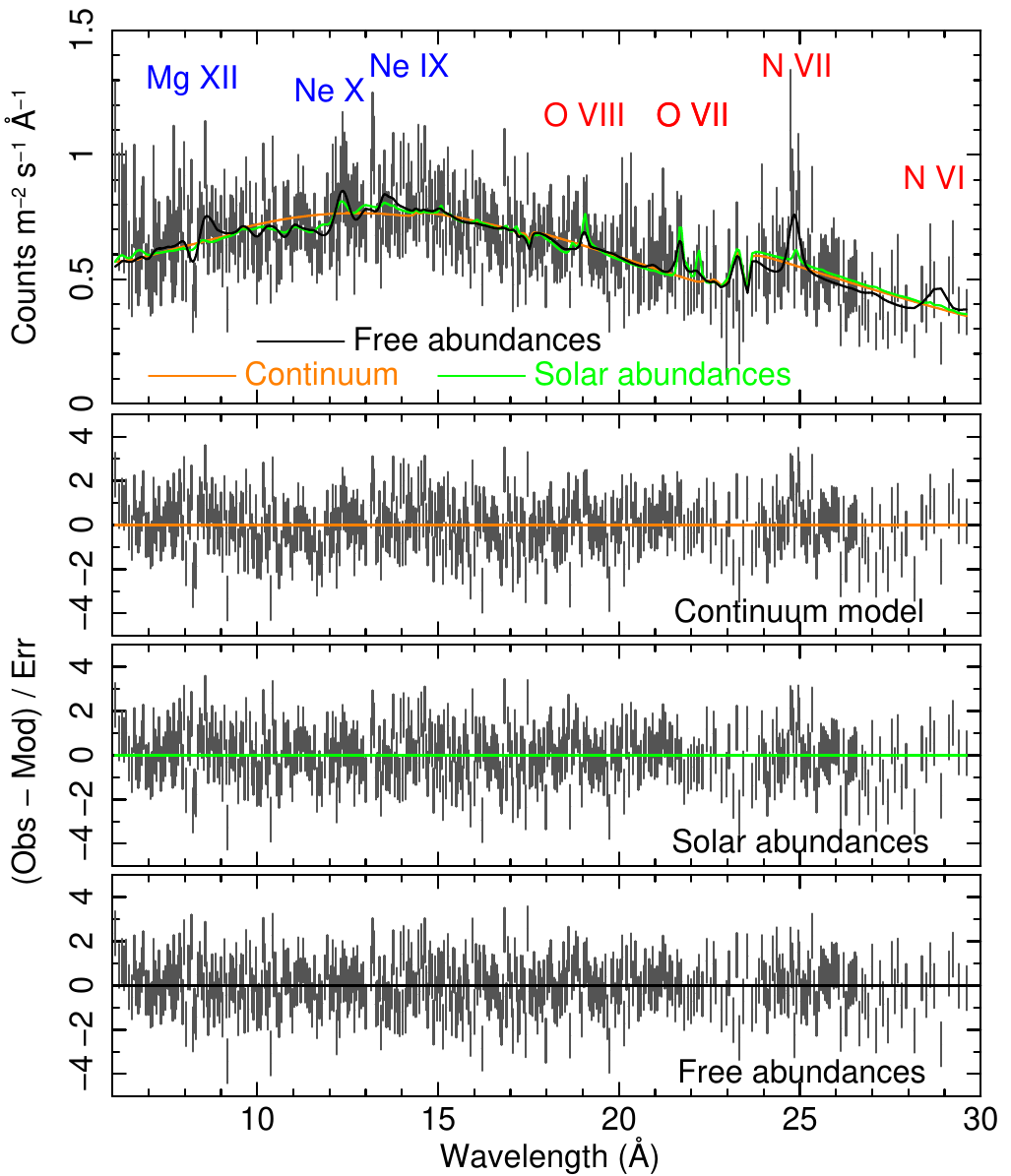}
		\caption{\small{From top to bottom: 2021 time averaged RGS spectrum and residuals for fits with the RHMM continuum-only model, the photoionised wind with Solar abundances and with free abundances. Rest-frame centroids of the typical, dominant H- / He-like transitions are labelled and colour-coded according to the contribution from the hot (blue) and cool (red) phases. Note the drop below 12\,{\AA} and the strong N\,{\scriptsize VII} line. }}
  
		\label{fig: RGS spectra}
		\end{figure}

The RGS and EPIC time-averaged spectra are shown in Fig. \ref{fig: RGS spectra}; they exhibit a very strong, narrow, emission feature around 24\,{\AA} close to the dominant 1s-2p transition of the N\,{\scriptsize VII}. There are additional features between 12-13\,{\AA} compatible with Ne\,{\scriptsize IX-X} transitions and a drop below 12\,{\AA} similar to that  detected in the ULXs NGC 1313 X-1 and NGC 5408 X-1 \citep{Pinto_2016}.

In previous work, the EPIC data in the 0.4-1.77 keV band were ignored to avoid smearing the narrow features and unnecessarily boosting model degeneracy owing to worse energy resolution for the EPIC (see e.g. \citealt{Pinto_2021}). 
We adopted the same approach and fitted the EPIC and RGS stacked spectra with the RHMM model. In all fits, the parameters of the spectral components and the ISM absorber were coupled among the EPIC and RGS models. However, we left the overall normalisations of the MOS 1,2 and RGS models free to vary with respect to the PN to account for the typical $\lesssim$ 5\% cross-calibration uncertainties. The RGS + EPIC continuum model provides an overall $C$-stat of 844 (and a comparable $\chi^2$) for 725 degrees of freedom (the detailed RGS spectrum is shown in Fig. \ref{fig: RGS spectra}). 

\subsection{Gaussian line scan}
\label{Gaussian line scan}

In order to search for and identify the strongest lines, we performed a standard scan of the spectra with a moving Gaussian line following the approach used in \citet{Pinto_2016}. We adopted a logarithmic grid of 1000 points
with energies between 0.4 (30\,{\AA}) and 10 keV (1.24\,{\AA}).
This provided an energy step that is comparable to the RGS and EPIC resolving power in their energy range. 
We tested line widths ranging from 100 to 10,000 km/s. 
At each energy we recorded the $\Delta C$ improvement to the best-fit continuum model and measured the single-trial significance as the square root of the $\Delta C$. To easily distinguish between emission and absorption lines, we multiplied the $\sqrt{\Delta C}$ by the sign of the Gaussian normalisation.

\begin{figure}
		\centering		\includegraphics[width=0.48\textwidth]{ 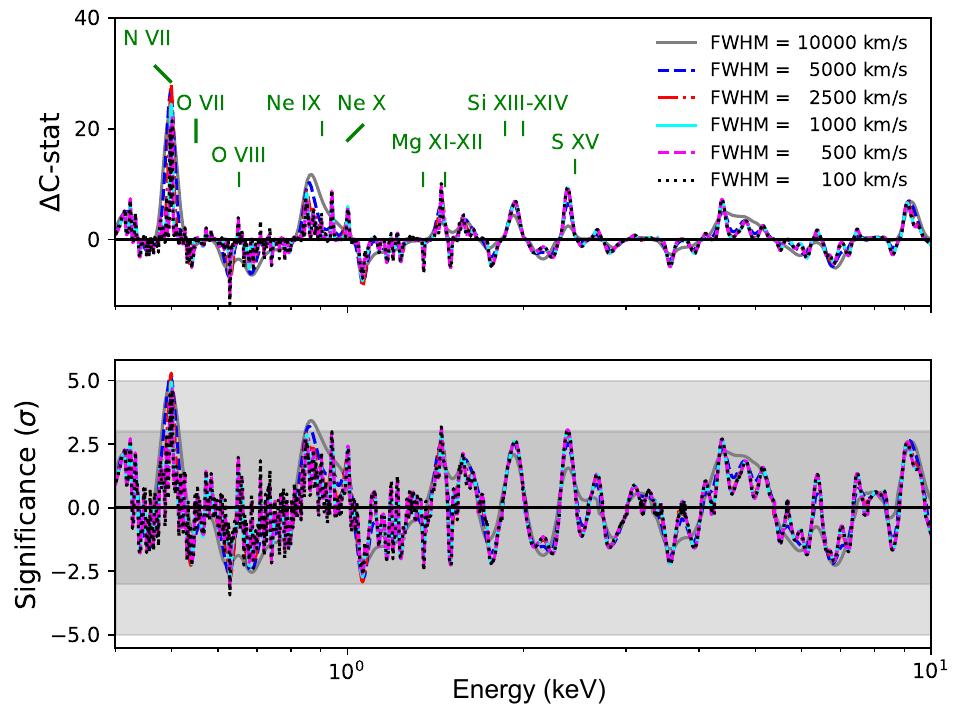}
        \vspace{-0.35cm}
		\caption{\small {\textcolor{black}{{Gaussian line scan for the 2021 time-averaged RGS (0.4-2 keV) and EPIC (2-10 keV) spectra and different line widths.}}}}
		\label{fig: RGS Gaussian line scan}
		\end{figure}

The results of the line scan obtained for the 2021 time-averaged stacked RGS + EPIC spectra are shown in Fig.\,\ref{fig: RGS Gaussian line scan}. We highlight that in the $0.4-2$ keV energy band ($7-30$\,{\AA}) only the RGS data is considered. The line scan picked out the strong emission-like feature near 0.5 keV (N\,{\scriptsize VII}) and a weaker one at 0.9 keV (Ne\,{\scriptsize IX}) as well as some faint absorption-like features between $0.6-0.8$ keV and above 1 keV. This shows a very good agreement with the CCD-quality EPIC spectra and supports our former approach of using three gaussian lines (see Fig. \ref{fig: best fits model for the obs. 0200470101} and Table \ref{table: XMM 0200470101 spectral fits}). The results of the Gaussian line scan are very similar to those obtained for NGC 1313 X-1 \citep{Pinto_2016} but the striking, close to rest-frame, N\,{\scriptsize VII} emission line would indicate a nitrogen-rich material which we discuss later.

\textcolor{black}{The single-trial line significance of the strongest feature from N\,{\scriptsize VII} is around 5\,$\sigma$, which is very high. Accounting for the look-elsewhere effect \textcolor{black}{and multiplying for the number of free bins within the whole 6--30\,{\AA} band (which would however not apply here since the line is close to rest) we can exclude at 3.5\,$\sigma$ that the feature is due to the noise (\citealt{Pinto_2016})}. The absorption features, if real, are offset from the dominant 1s-2p transitions and, therefore, have weaker significance \textcolor{black}{as previously reported for the archival data (\citealt{Kosec_2018a,Kosec_2021,Pinto_2020a})}.}

\subsection{Grids of photoionisation model}
\label{Grids of photoionisation models}

\textcolor{black}{Previous work  performed multi-dimensional scans of plasma models to fit multiple lines simultaneously, combining their individual improvements with respect to the continuum model (see, e.g., \citealt{Kosec_2018a}, \citealt{Pinto_2020b}). This technique has the advantage of a) preventing the fits from getting stuck in local minima and b) searching for multiple components or phases. Here we tried the same approach with the aim of locating velocity shifts in the emission and absorption lines. In order to avoid strong degeneracy in the plasma models, we adopted Solar abundances; this will have an impact on the overall spectral improvement due to the obvious nitrogen over-abundance, however, we address this when directly fitting the RGS spectra (see Sect. \ref{Photoionised and hybrid plasma modelling}).}

\textcolor{black}{The broadband spectral energy distribution (SED) is estimated by extrapolating the best-fit continuum model between 0.001 - 100 keV. We then computed the (photo-) ionisation balance with the {\scriptsize{SPEX}} component \textit{pion}  parameterised by the standard ionisation parameter, $\xi = L_{\rm ion} / ({n_{\rm H} \, R^2})$. Here, $L_{ion}$ is the ionising luminosity (measured between 13.6 eV and 13.6 keV), $n_{\rm H}$ is the hydrogen number density and $R$ is the distance from the ionising source. We also computed the stability (or $S$) curve, i.e. the relationship between the temperature and the ratio between the radiation and the thermal pressure, $\Xi = F / (n_{\rm H} c kT) = 19222 \, \xi / T$. 
The SED, ionisation balance $T - \xi$ and $S$ curves are shown in Fig. \ref{fig: ionisation balance}. 
The branches of the $S$ curve with a positive (negative) gradient are characterised by thermally stable (unstable) gas. Further technical details are provided in Fig. \ref{fig: ionisation balance}.}

\textcolor{black}{Following \citet{Pinto_2021}, we created grids of emission and absorption models using \textit{pion} and \textit{xabs} components in {\scriptsize{SPEX}}, respectively. For the line-emitting gas, we performed a multi-dimensional scan by adopting a logarithmic grid of ionisation parameters (log\,$\xi$ between 0 and 5 with steps of 0.1) and line-of-sight velocities, $v_{\rm LOS}$, between $-0.1c$ (blueshift) and $+0.1c$ (redshift, with steps of 500 km/s). We adopted a velocity dispersion, $v_{\sigma}$, of 1000 km/s which is the typical width of the features found in the Gaussian line scan (see Fig. \ref{fig: RGS Gaussian line scan}). As mentioned above, abundances were chosen to be Solar which saves a great deal of computing time. For the grids of absorption line models we chose a $v_{\rm LOS}$ range up to $-0.3c$ as done in previous work. 
The only free parameter for the \textit{pion} and \textit{xabs} components is the column density, $N_{\rm H}$.}

\textcolor{black}{We applied the automated routine scanning through the emission and absorption model grids fitting the Ho II X-1 2021 time-averaged EPIC and RGS spectra, again  using the RGS only in the 0.4$-$1.77 keV band. The results are shown in Fig. \ref{fig: physical model scan} in the 2D parameter space of line-of-sight velocity and ionisation parameter. The colour is coded according to the improvement relative to the continuum-only fit. Negative velocities refer to blueshifts or motion towards the observer. The emission model grids hint at a multi-phase gas. The solution with log $\xi$ = 0 and $v_{\rm LOS} = -0.018c$ is likely an artefact as it would require very low ionisation species moving at high velocities. The solutions that yield quasi-rest-frame emission lines are more feasible and would indicate the presence of cool (O\,{\scriptsize VII} - N\,{\scriptsize VII}) and hot (O\,{\scriptsize VIII} - Ne\,{\scriptsize IX}) gas describing the emission-like features shown in Fig. \ref{fig: RGS Gaussian line scan}. The absorption model grids favour a solution of plasma outflowing at $0.03c$ with a high ionisation state (log $\xi$ = 3.2), which is similar to the hotter solution for the emission lines. An alternative solution for the absorption lines yields parameters very similar to NGC 1313 X-1 (log $\xi$ = 2.8 and $v_{\rm LOS} = -0.19c$) \textcolor{black}{and a former tentative detection shown in \cite{Pinto_2020a} for Holmberg II X-1 archival data (log $\xi$ = 3.3 and $v_{\rm LOS} = -0.20c$)}. As expected, the changes in the $C$-stat are not very large, most likely because all the emission grids underpredict the N\,{\scriptsize VII} emission line. Instead, for the absorption features, the low $C$-stat is likely due to both adopted abundances and the spectral stacking between different observations which smears out the wind features in the LOS. The latter would smooth the lines because they are expected to be variable over timescales of days (\citealt{Kosec_2018a,Kosec_2021,Pinto_2020b}).} \textcolor{black}{All observations were taken 3-5 days apart from each other, which prevents us from building 2-3 ad-hoc large stacks and the signal-to-noise ratio of each spectrum is not sufficient to search for lines.}

\subsection{Photoionised and hybrid plasma modelling}
\label{Photoionised and hybrid plasma modelling}

\textcolor{black}{In order to obtain tighter constraints on the plasma models, we proceeded to model the RGS + EPIC spectra by adding absorption and emission components of plasma in photoionisation equilibrium with spectral parameters starting from the solutions found in the scans (Fig. \ref{fig: physical model scan}), at first keeping Solar abundances.}

\textcolor{black}{The addition of a single photoionised absorbing \textit{xabs} component onto the continuum model confirmed the results from the scan with  $v_{\rm LOS}=-0.033\pm0.003c$, log\,$\xi \,{\rm{ [erg/s \ cm]}}=3.3\pm0.1$, $v_{\sigma}=1500\pm1000$ km/s and $N_{\rm H} = (1.1 \pm 0.4) \times 10^{22}$ cm$^{-2}$. The fit improvement is still rather small ($\Delta C=16$) for it to be a significant detection.}

\textcolor{black}{The addition of a  photoionised emitting \textit{pion} component also provides results consistent with the model scan. In particular, $v_{\rm LOS}=+0.006\pm0.002c$ and log\,$\xi \,{\rm{ [erg/s \ cm]}}=1.15\pm0.15$. However, the fit with Solar abundances was poor ($\Delta C=13$) since it over-predicted the O\,{\scriptsize VIII} 19\,{\AA} line and under-predicted the N\,{\scriptsize VII} 24.8\,{\AA}. A single \textit{pion} component is also not able to fit all features, particularly those from higher ionic species such as Ne\,{\scriptsize IX-X} around 12-13\,{\AA}. A better description of these features are achieved by adding another \textit{pion} component with log\,$\xi \,{\rm{ [erg/s \ cm]}}=2.9\pm0.2$ at the same velocities. This fit with two emitting and one absorbing components with Solar abundances is shown in Fig. \ref{fig: RGS spectra} and corresponds to an overall $\Delta C=34$ improvement with respect to the continuum model for 10 degrees of freedom, which is a rather small improvement ($C/d.o.f=811/715$).}

\subsubsection{Non-Solar abundances and alternative plasma models}
\label{Non Solar abundances}

\textcolor{black}{Finally, we attempted at fitting again the RGS and EPIC spectra by releasing some of the abundances of the photoionised gas but keeping them coupled among the emission and absorption components. However, due to the weakness of most spectral features, we couple some of the elemental abundances. In the gaussian line scan and, visually in the RGS spectrum, the N\,{\scriptsize VII}, Ne\,{\scriptsize IX-X} and Mg\,{\scriptsize XII} lines appear significantly stronger than those from O\,{\scriptsize VII-VIII}. We therefore fitted the spectra by coupling the N, Ne, and Mg abundances and releasing the iron (which follows different evolutionary scenarios than the other three). Oxygen is kept fixed to Solar as hydrogen does not produce any lines in the 0.3-10 keV energy range. This provided a best fit with a final $C/d.o.f=784/713$ (see Fig. \ref{fig: RGS spectra}, top and bottom panel). This corresponds to a large improvement with respect to the continuum, $\Delta C=60$ (with $\Delta C$ = 22, 11, and 27 from the \textit{xabs}, the hot and cool \textit{pion} components respectively). The contribution of each component is therefore increased over the continuum when releasing the abundances despite being coupled with each other. There is substantial co-dependence between the \textit{xabs} (stronger) and the \textit{pion} (weaker) components. We obtained the following abundance ratios: O/H=1 (fixed), Fe/H=3, C-N-Ne-Mg/H=14 with uncertainties of $\sim50$\,\%.}

\textcolor{black}{In principle, the emission lines could be produced by plasmas in different conditions such as shocks within the ULX wind (especially for the higher ionisation component which may be supersonic) or photoionisation balance. The features can indeed be described with nearly indistinguishable fits and comparable $C$-stat values by substituting the two \textit{pion} components with two collisionally-ionised emission components (\textit{cie} in {\scriptsize{SPEX}}, with temperatures k$T_{\rm cool}\sim 0.17$ and k$T_{\rm hot}\sim 1.2$ keV \textcolor{black}{and the abundance ratios being consistent with those obtained with the two \textit{pion} components}).}

\section{Discussion}
\label{Discussion}
This paper focuses on the physical processes that might cause spectral transitions in Ho II X-1 and in general in ULXs. \textcolor{black}{We first discuss the broadband properties of the XMM-\textit{Newton} and NuSTAR spectra, and the possible presence of a hard tail above 10\,keV. Then we search for a correlation between the total luminosity, the temperature and the radius of the thermal components. Finally, we compare our results with those obtained for different sources and discuss the properties of the wind.}


\subsection{Broadband spectral properties}
\label{X-ray broadband spectroscopy}
Ho II X-1 shows a flux variability of a factor 2-3 within a given observation and a factor up to 6-7 in different observations, as shown in Fig. \ref{fig: lightcurves} (left-top panel). In addition there are a few flux drops  (the phenomenon is more evident in long-term monitoring with \textit{Swift}/XRT, see e.g \citealt{Gurpide_2021b}) of the order of few ks due probably to wind optically-thick clumps that briefly obscure the innermost part of the disc, with a reduction of the spectral hardness (Fig. \ref{fig: lightcurves}, lower-left panel). This can be caused either by an increase in the accretion rate or by the  precession of the disc, although the lack of periodicity in the phenomenon would favour the former one.
Our results showed that the spectrum is well described with a double thermal model (two \textit{mbb} components), in agreement with the general properties of intermediate-soft ULX spectra (\citealt{Sutton_2013}). Alternative models using comptonisation for the hotter component provides comparably good fits but the parameters are unconstrained for low-statistics short observations and, therefore, we focussed primarily on the double thermal model. 

We also tested a model with a third component (a cutoff powerlaw) describing the hard tail in the fashion of a NS magnetosphere and obtained similar results. However, we note that it mainly affects the band beyond 10 keV (see, e.g., \citealt{Walton_2015}) and, therefore, the powerlaw parameters could not be well constrained. 
\textcolor{black}{In order to determine the role of such third component, we also fit the XMM-Newton data (obsid: 0724810101-301) with the simultaneous NuSTAR data (obsid: 30001031002/3, 30001031005) (see Fig. \ref{fig: XMM+NuSTAR modeling with RHMMP}), which were retrieved from \citet{Walton_2015}. These data were fitted with the RHMM model with a cutoff powerlaw  ($E_{\rm cutoff}$= 7.9 keV and $\Gamma=0.59$, see \citealt{Walton_2020}). The cool and hot component are characterized by a temperature $\sim$ 0.3 keV and $\sim$ 1.8 keV, respectively. As expected, the addition of a third component (\textit{pow}+\textit{etau}), does not dramatically influence the constraints of the two \textit{mbb} components.} 

\begin{figure}
		\centering
        \includegraphics[width=0.5\textwidth]{ 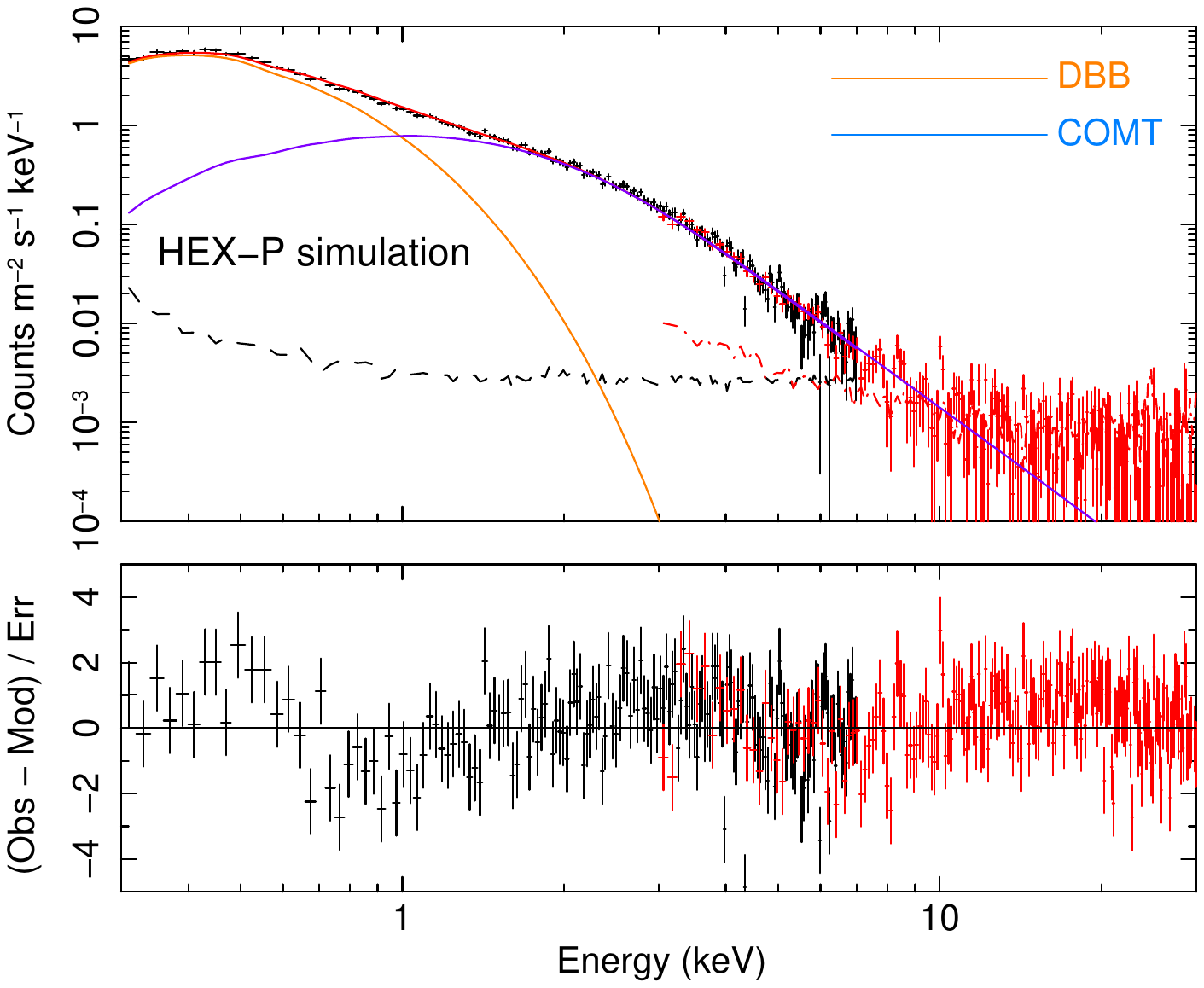}
		\caption{{\small HEX-P simulation with an exposure time of 500 ks. The simulated spectra were generated using the RHBDP model (with the parameters obtained from the best fit of the low flux observation ID 0864550201). The spectra were then fitted with the disc blackbody + comptonisation components (RHDCom model, ${\chi}^2$/ d.o.f = 485/356). The disc blackbody and the comptonisation components are represented in orange and blue, respectively. The residuals show that a third component is necessary to describe the simulated spectra above 8 keV. }}
		\label{fig:HEX-P simulation}
        \vspace{-0.3cm}
		\end{figure}

\textcolor{black}{The data collected by NuSTAR  above 7 keV are overwhelmed by the background when the source is at low-flux. Constraints on the hard energy tail above 10 keV are very hard to set. A constraint on the properties of such hard X-ray component in the low-flux regime is necessary to understand its nature and help distinguish between a magnetosphere typical of a NS and a small corona around a BH. In order to study this, we proceed to perform a low-flux simulations of Ho II X-1 spectra with the HEX-P satellite (see Fig. \ref{fig:HEX-P simulation}, response matrices from July 2023; \citealt{Stern_2023}). HEX-P will be able to detect the source at 10\,keV and above, which is necessary to place constraints on the tail and the inner part of the disc. This may unveil the presence of an accretion column and the nature of the compact object.}

The thermal components appear hotter at higher luminosity (see Table \ref{table: Table results for the best fit RHMM model for all observations. }) maybe due to an increase of the accretion rate $\dot{M}$ or a better view of the inner accretion flow (wind photosphere less dense due to a precessing disc, \textcolor{black}{although this should not be the case for the cooler, outer, component}). 


\subsection{Luminosity - Temperature relations}
\begin{figure}
		\centering
		\includegraphics[width=0.48\textwidth]{ 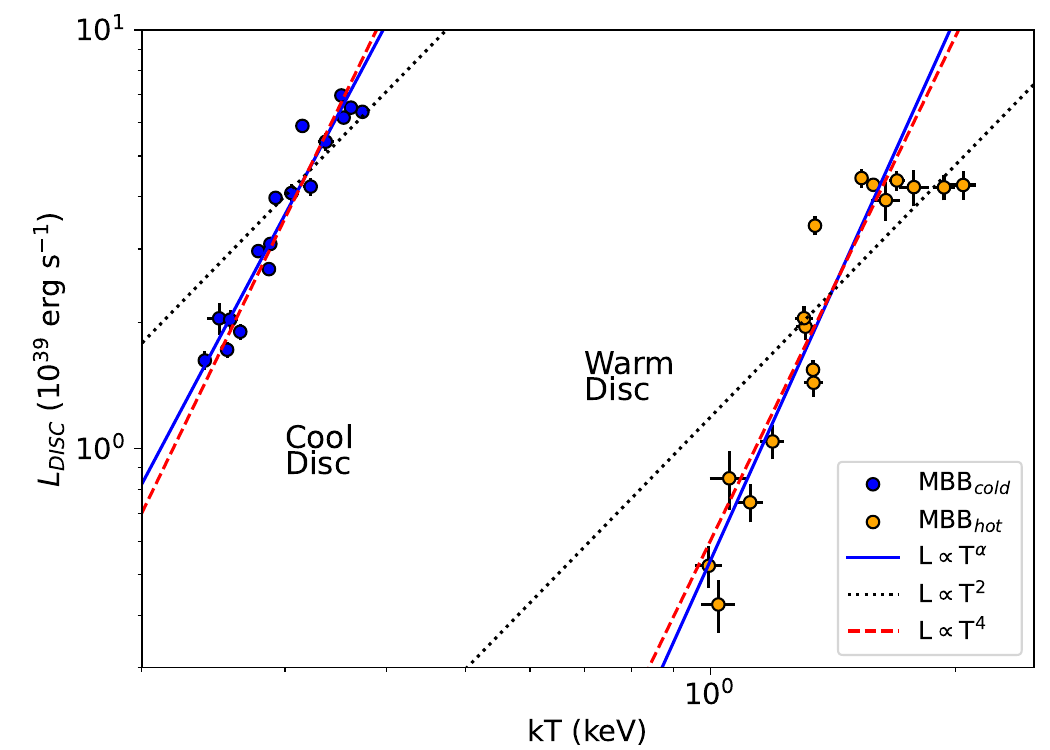}
		\caption{{\small Luminosity-Temperature plot for the RHMM model. Bolometric luminosity (0.001-1000 keV) versus temperature for the cool (in blue) and hot (in orange) components. The black solid, the black-dotted and the red dashed lines represent the regression line and the two theoretical models of the slim disc and thin disc models, respectively.}}
		\label{fig: L-T plot} 
		\end{figure}

In order to unveil any disc structural changes with the time, we examined the luminosity vs temperature (L-T) behaviour for both thermal components of the best fit RHMM model. In particular, it is useful to compare the L-T trends with the theoretical expectations for a thin disc model with a constant emitting area (L $\propto$ $T^{4}$, \citealt{SS1973}) and advection dominated model (L $\propto$ $T^{2}$,\citealt{10.1093/pasj/53.5.915}).
For each thermal component we estimate the bolometric luminosity extrapolating down to 0.001 and up 1000 keV from the best-fit models. There is a small uncertainty in the total luminosity owing to our ignorance of the UV flux, but we only expect those to make a minor contribution for such a bright intermediate-hardness ULX (see discussion on the spectral energy distribution in \citealt{Pinto_2023}).
The luminosity-temperature results are reported in Figure \ref{fig: L-T plot} and Table \ref{table:pearsons and spearman correlation coefficients}, which clearly show positive correlations between luminosity and temperature for both components. In addition to the RHMM model, we also explore the results for the RHBDP model in Appendix \ref{additional figures and tables}. \textcolor{black}{We find a general agreement although there are some deviations with respect to the RHMM model for the hot thermal component at high luminosities.}

The L-T plot confirms that the source becomes spectrally harder when brighter as seen in Fig. \ref{fig: lightcurves} (lower-left panel).
In both plots and for each thermal component we compute the regression lines of the L-T points. We also quantify the L-T trends by computing the Pearsons and Sperman correlation coefficients by using \textit{scipy.stats.pearsonsr} and \textit{scipy.stats.spearmanr} routines in {\scriptsize{PYTHON}}. These are reported in Table \ref{table:pearsons and spearman correlation coefficients}; for both thermal component, a strong correlation is suggested. At low luminosities, the L-T trends of all models broadly agree with a $L \propto T^4$ relationship as predicted for Eddington-limited thin discs (\citealt{SS1973}). \textcolor{black}{There are significant deviations at high luminosities, especially for the \textcolor{black}{hot} component which exhibit a large scatter from a regression line (see Fig. \ref{fig: L-T plot}). The exact trend depends on the model adopted. For the RHBDP model (Fig. \ref{fig:L-T plots for alternative models}, middle panel) the temperatures are lower than predicted for a thin disc with constant emitting area, suggesting a possible expansion of the disc and/or a contribution to the emission from the wind. Instead, for the RHMM model the scatter is less clear although it seems to saturate} (see    
Fig. \ref{fig: L-T plot}). These deviations appear for a total bolometric luminosity greater than 5 $\times \  10^{39} \  \rm{erg/s}$. As seen for NGC 55 ULX-1 (\citealt{Barra_2022}), deviations from sub-Eddington behaviour could potentially be used to place rough constraints on the mass of the compact object. If we assume the deviations are due to the fact the source is \textcolor{black}{exceeding} the Eddington limit $L_{\rm Edd} = 1.4 \times 10^{38} \frac{M}{M\textsubscript{\(\odot\)} }$ erg/s or the critical luminosity $L_{\rm critical}$= $\frac{9}{4} L_{\rm Edd}$ (\citealt{Poutanen_2007}), we estimate a mass of the compact object ranging between 16 and 36 $M\textsubscript{\(\odot\)}$. This would suggest a stellar mass black hole in agreement with estimates inferred from the powerful radio jets from this source (see \citealt{Cseh_2014}).  However, deviations were found in Galactic XRBs when the source exceeded 30\% of the Eddington luminosity (e.g. \citealt{Steiner_2009,Lancova_2019}) which might indicate a slightly more massive BH. \textcolor{black}{Given that Holmberg II X-1 always shows the typical characteristics of a ULX (soft spectra, turnover, high luminosity and lines) it is more likely that the accretion rate is around Eddington and, during these deviations, exceeds the critical value, altering the structure of the accretion disc.}

\begin{center}
	\begin{table}
	\caption{Pearsons/Spearman coefficients and slope from the regression least square for the L-T trends, for both cool and hot components, by using the best fit values of the RHMM model.}  
	 \renewcommand{\arraystretch}{1.}
 \small\addtolength{\tabcolsep}{0pt}
 \vspace{0.3cm}
	\centering
	\scalebox{1}{%
  \begin{tabular}{c c c}
    \hline
    \multirow{2}{*}{Correlation coefficient} &
      \multicolumn{2}{c}{RHMM}  \\ \cline{2-3}
    &  ${(L-T)}_{cool}$ & ${(L-T)}_{hot}$   \\
    \hline
    Pearsons & 0.96 & 0.90  \\
    \hline
    Spearman & 0.95 & 0.87  \\
    \hline
    Slope & 3.66 $\pm$ 0.29 & 4.30 $\pm$ 0.63 \\
    \hline
  \end{tabular}}   \label{table:pearsons and spearman correlation coefficients}
  \begin{quotation}
\end{quotation}
        \vspace{-0.3cm}
\end{table}
\end{center}

\subsection{Radius - Temperature relation}
\begin{figure}
		\centering
		\renewcommand{\arraystretch}{1.}


%
        \includegraphics[width=0.475\textwidth]{ 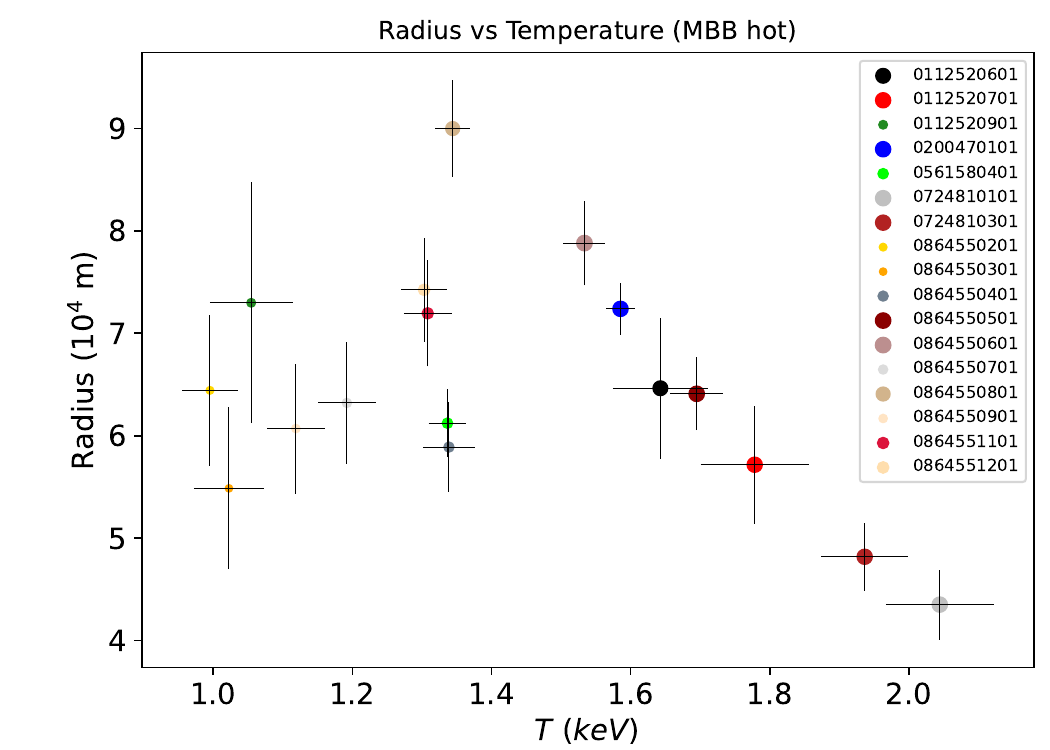}
        \includegraphics[width=0.475\textwidth]{ 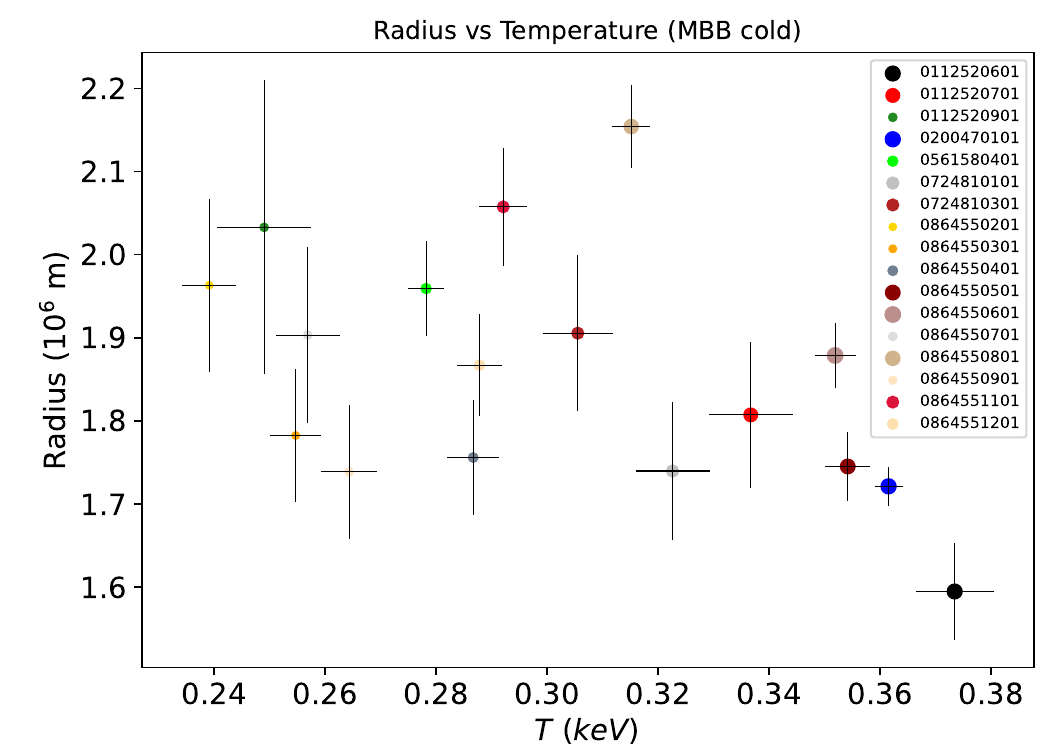}
		\caption{{\small Radius-Temperature plot for the thermal components of the RHMM model. The coloured circles scale with the luminosity of the source in each specific observation.}}
		\label{fig: R-T plot 2}
        \vspace{-0.3cm}
		\end{figure}

We show the radius-temperature (R-T) plots for the RHMM model in Fig. \ref{fig: R-T plot 2}; \textcolor{black}{we do not present the results for the other models because they are consistent within the uncertainties.  The radii are estimated from the relation between the luminosity and the temperature of a blackbody.} In all cases, there is no clear trend between radii of the two thermal components and their temperatures. The radii would appear constant \textcolor{black}{at confidence levels within 90-95\,\%}. 
The radius of the hot component has a value ranging between 50-90 km; instead the radius of the cold component is ranging in the range 1500-2200 km. \textcolor{black}{The tests performed with the alternative models yield a similar order of magnitude for such radii.} By assuming an average value of the mass of the compact object of 25 M$_{\odot}$ (from the range obtained through the Luminosity-Temperature relation), these correspond to, respectively, 2 $R_{G}$ and 80 $R_{G}$. The values found might refer to the inner and the outer part of the disc, respectively. Indeed, \textcolor{black}{for $\dot{M} \sim \dot{M}_{\rm Critical} = 9/4\dot{M}_{\rm Edd}$ we would obtain a spherisation radius $R_{\rm sph}= \frac{27}{4} \frac{\dot{M}}{\dot{M}_{\rm Edd}} \, R_{G} \, \sim 15 R_{G}$.}

We note that in the case of a NS or very small BH accretor (up to a few times $M\textsubscript{\(\odot\)}$) the rate would be highly super-Eddington, \textcolor{black}{$R_{\rm sph}$ would be larger by a factor of a few,} and the models might fail to reproduce the exact values of disc radii and temperature, although our L-T trends would disfavour it.
 
\


\subsection{Comparison with other ULXs}
\label{sec:comparison_with_other_ulxs}

A similar study was performed for the soft ultraluminous X-ray source NGC 55 ULX-1 (\citealt{Barra_2022}). Also in this case, the regression line in the L-T plot agreed with the thin disc and the deviations implied a 6-14 $M\textsubscript{\(\odot\)}$ stellar-mass black hole and smaller $\dot{M}$. In the case of NGC 1313 X-1, \textcolor{black}{deviations from a single L-T track} are present for the hot component linked to an obscuring wind or a change of the disc scale height, while there is no L-T correlation for the cool component, maybe due to the modeling of hotter part of the accretion disc (see, e.g., \citealt{Walton_2020}, \citealt{Miller_2013}; although see \citealt{Gurpide_2021a} for the caveats and related to the modelling and the associated trends).

Ho II X-1 is similar to NGC 5204 X-1 regarding repeated transitions between the three classical ULXs regimes (HUL, SUL and SSUL/ULS, \citealt{Gurpide_2021b}). However, according to their long-term \textit{Swift}/XRT light curves, Ho II X-1 remains around a 0.3-10 keV count rate of 0.15-0.20 c/s for most of the time with an isolated long episode with the count rate switching between 0.05 and 0.30 c/s, which correspond to SUL and SSUL regimes, respectively.

These transitions may be driven by geometrical effects (the funnel structure) and wind due to a varying accretion rate and/or precession, and a line-of-sight grazing the wind  allowing for a rapid transition between the SUL-SSUL regimes. These transitions were also found during the epochs of highest flux of the ULXs NGC 55 ULX-1 and NGC 247 ULX-1 (see e.g. \citealt{Stobbart_2004, D'Ai_2021}) although for these two ULXs, and especially the latter, the flux drops are more prominent. This would therefore suggest a lower inclination angle in Ho II X-1  (for a comparable $\dot{M}$) or a lower accretion rate and, therefore, a thinner wind along the line of sight.

\subsection{Wind properties}
\label{sec:Wind properties}


\textcolor{black}{X-ray spectra from CCD / imaging detectors such as EPIC have lower spectral resolution than gratings but provide a far greater number of counts and may therefore yield important results provided that there are some dominant and reasonably well isolated lines. This is the case for the feature at $0.5$ keV from N\,{\scriptsize VII} (which is well resolved in the RGS stacked spectrum, see Fig. \ref{fig: RGS spectra}). Its flux \textcolor{black}{weakly} correlates with the source bolometric luminosity, see Fig. 
\ref{fig: Normalization_vs_Bolometric Luminosity plot}. This line is produced along with others by a cool plasma (see Fig. \ref{fig: RGS spectra}) \textcolor{black}{perhaps associated with the cool thermal component}. Should such trend be confirmed with deep flux-resolved RGS spectra it would suggest an increase of the accretion rate (as the L-T trends indicate) because a changing viewing angle or ionisation would produce an opposite trend. A parallel paper will also make use of NICER and \textit{Swift} X-ray observations and data from other energy bands to investigate the variability of the dominant spectral features on timescales from a few days to several weeks.}

\textcolor{black}{The physical modelling of the spectral lines present in the RGS data indicate evidence for a multiphase plasma. The phase responsible for the strong N\,{\scriptsize VII} emission line (log $\xi\sim1$) appears distinct from the two hotter absorption and emitting components (log $\xi\sim3$). Although the $T-\Xi$ curve appears generally stable to thermal perturbations, there is evidence for some possible points of instability which would indeed split the plasma into multiple phases (see Fig. \ref{fig: ionisation balance}). The two hot components are not independent of each other due to the low velocity of the absorbing gas; in other words, the inclusion of one weakens the significance of the other although the absorption is clearly stronger. This may also be due to the detailed shapes of the features which resemble those of P-Cygni profiles, especially that of Mg\,{\scriptsize XII} and Ne\,{\scriptsize X} at 8.4 and 12.1\,{\AA}, respectively (see Fig. \ref{fig: RGS spectra}). However, the statistics are not good enough to make actual claims of P-Cygni profiles although the physical modelling would confirm them through weakly-redshifted emission and blue-shifted absorption (see Fig. \ref{fig: physical model scan}). Due to this mutual dependence, the fit of the abundances (but also the strength of the N\,{\scriptsize VII} line) makes performing Monte Carlo simulations for the physical models (as done in previous work, e.g. \citealt{Kosec_2021,Pinto_2021}) challenging because degeneracy prevents a meaningful parameter space exploration. The fact that the spectral improvement of each component is boosted when releasing the abundances despite being coupled with each other would suggest that they are all present. Indeed, \citet{Goad_2006} first report on the presence of a possible emission excess from the O\,{\scriptsize VII} triplet around 22\,{\AA} in the deepest, individual, RGS spectrum of Ho II X-1 from the archival observation taken in 2004.
Moreover, \citet{Kosec_2018a} and, later on, \citet{Kosec_2021} studied these data as well as the 2002-2013 time average RGS spectrum (137\,ks) and found features consistent with those reported in this work. The 2021 time averaged RGS spectrum seems to show variability with respect to the archival data \citep{Kosec_2018a,Kosec_2021}, in which the dominant feature was the O\,{\scriptsize VII} line due probably to a variation in the ionisation parameter and so in the luminosity. }

The line-emitting plasma components are consistent with being at rest within 2-3\,$\sigma$, each with a velocity dispersion of about 1000-3000 km/s which is very similar (albeit cooler or with lower ionisation parameters) to that observed in the ubiquitous winds of Eddington-limited Galactic X-ray binaries (for a review, see \citealt{Neilsen_2023}). This might suggest a common mechanism such as thermal driving \citep{Middleton_2022}. From simultaneous fits of the He-like triplets of Mg\,{\scriptsize XI}, Ne\,{\scriptsize IX} and especially O\,{\scriptsize VII} we obtain a 2\,$\sigma$ upper limit on the gas density of about $1.5\times10^{11}$ cm$^{-3}$ (with the most probable value of $10^{10}$ cm$^{-3}$) in agreement with results by \cite{Pinto_2020a} for the ULX NGC 1313 X-1 and for line-emitting plasma of Galactic XRBs \citep{Psaradaki_2018}, which reinforces our hypothesis for a common mechanism. This was expected given the very weak, undetected, O\,{\scriptsize VII} forbidden line at 22.1\,{\AA} \textcolor{black}{with respect to the unresolved resonance and intercombination line complex at 21.6-21.8\,{\AA} seen both in the spectrum and the line scan (Fig. \ref{fig: RGS spectra} and \ref{fig: RGS Gaussian line scan}).} Of course, the wind in Ho II X-1 is more extreme than in XRBs given that each line-emitting components has an X-ray luminosity, $L_{\rm [0.3-10 \ keV]}$, of about $10^{38}$ erg/s, i.e. orders of magnitude higher than in Galactic XRB winds.

The super-Solar abundance of nitrogen in the wind of Holmberg II X-1 is most likely a signature of the abundance pattern of the donor companion star. This could provide alternative means for constraining its nature because previous work has shown that the optical and infrared fluxes are strongly affected by the irradiated outer disc and the surrounding nebula, making it difficult to retrieve information on the companion star \citep{Tao_2012,Heida_2016,Lau_2019}. N-rich  stars with a super-Solar N/Fe ratio are found in low-metallicity environments. Some examples are B-type giant/supergiant stars, \textcolor{black}{ as reported for Ho II X-1 where the optical counterpart is consistent with a O/B type companion star (\citealt{Kaaret_2004b})}, or Wolf–Rayet stars in young star clusters of the Magellanic clouds and similar dwarf galaxies \citep{Lennon_2003,Yarovova_2023}. In some cases they might be luminous evolved stars on the asymptotic giant branch (AGB) and more regular red giant stars in the LMC. The latter have been interpreted as the observational evidence
of the result of tidally shredded stellar clusters (see \citealt{Fernandez_2020}). This is supported in the case of Holmberg II X-1 by the discovery of a bow shock which indicates that the ULX binary system was likely ejected from the nearby young star cluster \citep{Egorov_2017}. \textcolor{black}{These results suggest that Holmberg II X-1 system was born in a low-metallicity environment, rich in young/massive stars, and escaped from the native stellar cluster.}


\section{Conclusions}
\label{Conclusions}
In this paper we have performed an X-ray spectral analysis to understand the nature of the ultraluminous X-ray source Holmberg II X-1. Two thermal components provide a good description of the 0.3-10 keV spectra with the hotter component broadened perhaps by Compton scattering.
Both thermal components become hotter at higher luminosities indicating an increase in the accretion rate (most likely) or a clearer view of the inner part of the accretion flow due to a disc precession.
The trends between the bolometric luminosity and temperature of each component broadly agree, at low luminosities, with the L $\propto$ $T^{4}$ relationship expected from Eddington-limited thin disc. 
Deviations are seen above $5 \times 10^{39} \rm erg/s$ which are likely due to the accretion rate exceeding the supercritical rate; this would imply a black hole with a mass ranging between 16-36 $M\textsubscript{\(\odot\)}$ as the compact object.
Our results suggest that the two thermal components unveil the portions of the disc inside (hotter) and outside (cooler) the spherisation radius where the wind is launched, in agreement with the super-Eddington scenario. The wind abundance pattern strongly favours a nitrogen-rich donor star, likely a supergiant, which has escaped from its native stellar cluster \textcolor{black}{characterised by} a low-metallicity environment.

\section*{Acknowledgements}

This work is based on observations obtained with XMM-\textit{Newton}, an ESA science mission funded by ESA Member States and the USA (NASA). CP acknowledges support for PRIN MUR 2022 SEAWIND 2022Y2T94C and INAF LG 2023 BLOSSOM.
TDS acknowledges support from PRIN-INAF 2019 with the project "Probing the geometry of accretion: from theory to observations" (PI: Belloni).
TPR acknowledges funding from STFC as part of the consolidated grants ST/T000244/1 and ST/X001075/1.
\section*{Data Availability}

All the data and software used in this work are publicly available from ESA's XMM-\textit{Newton} science Archive (XSA\footnote{https://www.cosmos.esa.int/web/XMM-Newton/xsa}) and NASA's HEASARC archive\footnote{https://heasarc.gsfc.nasa.gov/}. Our spectral codes and automated scanning routines are publicly available and can be found on the GitHub\footnote{https://github.com/ciropinto1982}.



\bibliographystyle{aa}
\bibliography{bibliography} 



\clearpage


\appendix

\section{Additional figures and tables}
\label{additional figures and tables} 




	\begin{table*}
\caption{Results of the RHBDP model for all observations.}  
 \renewcommand{\arraystretch}{1.}
 \small\addtolength{\tabcolsep}{0pt}
 \vspace{0.1cm}
	\centering
	\scalebox{.83}{%
	\begin{tabular}{cccccccccccc}
    \toprule
    {{Obs. ID}}  &
    {{$\rm{N_{H}}$}}  &
    {$\rm{kT_{bb}}$} [keV] &
    {$\rm{kT_{dbb}}$} [keV] &
    {$\rm{Norm_{pow}}$} & 
    {$\rm{L_{X\,bb}}$} [erg/s] & 
    {$\rm{L_{X\,dbb}}$} [erg/s] & 
    {$\rm{L_{X\,pow}}$} [erg/s] & 
    ${\chi}^2$/ d.o.f &
    ${\chi}^2_{\rm{PN}}$ &
    ${\chi}^2_{\rm{MOS1}}$ &
    ${\chi}^2_{\rm{MOS2}}$ \\
    
    \midrule
0112520601 & 0.51 $\pm$ $_{0.01}^{0.12}$ & 0.187 $\pm$ 0.005 & 1.35 $\pm$ 0.06 & 1484.1 $\pm$ 100.2 & 2.3 $\pm$            0.5 & 4.1 $\pm$      1.0 & 4.3 $\pm$   0.3  & 229/256 & 81 & 61  & 87  \\\midrule
0112520701 & 0.50 $\pm$ $_{0.0}^{0.03}$  & 0.185 $\pm$ 0.004 & 1.7 $\pm$ 0.1 & 1561.4 $\pm$ 168.1 & 2.5 $\pm$             0.2  & 3.2 $\pm$      1.0 & 4.5 $\pm$  0.5  & 228/216 & 99 & 76  & 53  \\\midrule
0112520901 & 0.50 $\pm$ $_{0.0}^{0.17}$  & 0.158 $\pm$ 0.004 & 1.6 $\pm$ 0.2 & 52.2 $\pm$ $_{52.2}^{57.5}$ & 1.0 $\pm$ 0.2 & 0.9 $\pm$  0.4 & 0.1 $\pm$  0.1     & 119/117 & 52 & 26 & 41  \\\midrule
0200470101 & 0.500 $\pm$ $_{0.00}^{0.004}$ & 0.194 $\pm$ 0.001 &  1.63 $\pm$ 0.03  & 1161.0 $\pm$ 40.6  & 2.89 $\pm$ 0.06   & 4.2 $\pm$  0.3 &   3.4 $\pm$   0.1    & 622/403 & 222 & 180 & 220 \\\midrule
0561580401 & 0.50 $\pm$ $_{0.00}^{0.01}$ & 0.179 $\pm$ 0.002 &  1.62 $\pm$ 0.05 & 330.5 $\pm$ 30.7  & 1.39 $\pm$     0.05  &  1.5 $\pm$ 0.2 & 0.96 $\pm$     0.09 & 347/291 & 160 & 73 & 114 \\\midrule
0724810101 & 0.54 $\pm$ $_{0.04}^{0.14}$ & 0.152 $\pm$ 0.006 &  1.8 $\pm$  0.2  & 1823.1 $\pm$ 163.0  & 2.0 $\pm$      0.4  &  2.4 $\pm$ 0.9 &  5.3 $\pm$     0.5   & 234/253 & 60 & 90 & 84 \\\midrule
0724810301 & 0.50 $\pm$ $_{0.0}^{0.02}$ & 0.172 $\pm$ 0.004  &  1.8 $\pm$  0.1      &  1649.0 $\pm$ 141.4 & 1.8 $\pm$ 0.1 &  2.5 $\pm$ 0.7  &  4.8 $\pm$    0.4   & 260/254 & 91 & 90 & 79 \\\midrule
0864550201 & 0.50 $\pm$ $_{0.0}^{0.04}$ & 0.192 $\pm$ 0.003 &  1.5 $\pm$  0.1     & 32.2 $\pm$ 25.4 & 0.76 $\pm$       0.06  &  0.6 $\pm$ 0.1 &   0.09 $\pm$ 0.07   & 198/167 & 74 & 65 & 59  \\\midrule
0864550301 & 0.65 $\pm$ $_{0.12}^{0.13}$ & 0.190 $\pm$ 0.006 &  1.21 $\pm$ 0.09 & 83.3 $\pm$ 18.7  & 0.9 $\pm$         0.2 &  0.5 $\pm$ 0.2 &  0.24 $\pm$  0.05     & 220/158  & 73 & 52 & 95 \\\midrule
0864550401 & 0.50 $\pm$ $_{0.0}^{0.01}$  & 0.152 $\pm$ 0.002 &  1.74 $\pm$ 0.09  & 227.3 $\pm$ 42.0  & 1.29 $\pm$     0.07  &  1.4 $\pm$ 0.3 &  0.7 $\pm$    0.1    & 312/224 & 128 & 92 & 100  \\\midrule
0864550501 & 0.52 $\pm$ $_{0.02}^{0.07}$  & 0.179 $\pm$ 0.003 &  1.68 $\pm$ 0.07 & 1397.3 $\pm$ 87.0  & 2.9 $\pm$      0.3  &  3.8 $\pm$ 0.6 &  4.0 $\pm$     0.2   &  326/326 & 90 & 119 & 127  \\\midrule
0864550601 & 0.50 $\pm$ $_{0.00}^{0.01}$ & 0.162 $\pm$ 0.002 &  1.56 $\pm$ 0.04  & 1255.0 $\pm$ 62.0  & 3.04 $\pm$   0.09  &  4.4 $\pm$ 0.5 &  3.6 $\pm$     0.2  & 413/342 & 143 & 147 & 123 \\\midrule
0864550701 & 0.50 $\pm$ $_{0.0}^{0.05}$  & 0.174 $\pm$ 0.003 &  1.40 $\pm$ 0.07  & 240.2 $\pm$ 29.3  & 0.87 $\pm$      0.09  &  1.0 $\pm$ 0.2 &   0.7 $\pm$  0.1  & 218/190  & 86 & 72 & 60  \\\midrule
0864550801 & 0.50 $\pm$ $_{0.0}^{0.01}$  & 0.168 $\pm$ 0.002 &  1.56 $\pm$ 0.04  & 728.7 $\pm$ 48.2  & 2.69 $\pm$      0.09  &  3.5 $\pm$ 0.4  &   2.1 $\pm$ 0.1  & 426/318  & 156 & 131 & 139  \\\midrule
0864550901 & 0.50 $\pm$ $_{0.0}^{0.05}$  & 0.171 $\pm$ 0.003 &  1.43 $\pm$ 0.08  & 136.4 $\pm$ 27.3  & 0.89 $\pm$      0.07  &  0.8 $\pm$ 0.2  &  0.4 $\pm$ 0.1   & 230/191  & 113 & 53 & 64  \\\midrule
0864551101 & 0.50 $\pm$ $_{0.0}^{0.01}$  & 0.154 $\pm$ 0.002 &  1.68 $\pm$ 0.07  & 308.6 $\pm$ 52.0  & 1.93 $\pm$      0.09  &  2.0 $\pm$ 0.3 & 0.90 $\pm$   0.1 & 316/247  & 132 & 87 & 97  \\\midrule
0864551201 & 0.50 $\pm$ $_{0.0}^{0.01}$ & 0.158 $\pm$ 0.002  &  1.51 $\pm$ 0.05  & 484.6 $\pm$ 43.9  &  1.95 $\pm$     0.09  &  2.1 $\pm$ 0.3 &  1.4 $\pm$   0.1 & 323/270  & 109 & 87 & 127  \\\midrule
    \bottomrule
    \end{tabular}}
    \label{table: Table results for the best fit RHBDP model for all observations. }
      \begin{tablenotes}
      \small 
  \item[] The units of the parameters are the same of Table \ref{table: XMM 0200470101 spectral fits}. The slope $\Gamma= 0.59$ and the optical depth $\tau_{0}= 0.1265$, at a corresponding $\rm{E_{cutoff}}$ = 7.9 keV, are fixed for all observations.  $\rm{Norm_{pow}}$ is expressed in $10^{44}$ ph/s/keV unit.
      \end{tablenotes}
      
\end{table*}




		\begin{figure}
		\centering
		\includegraphics[width=0.45\textwidth]{ 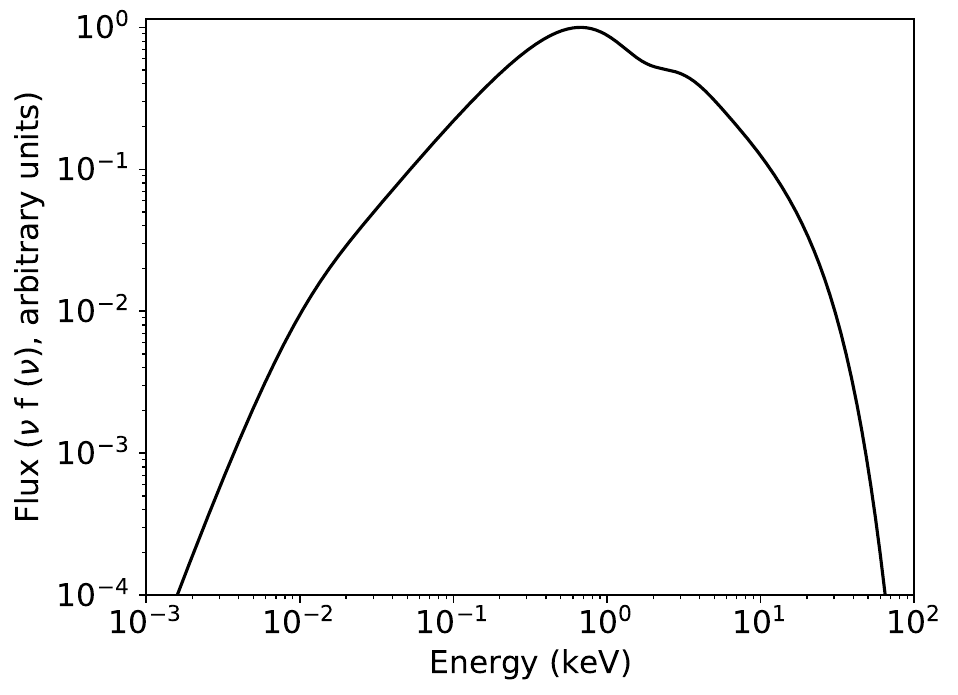}
		\includegraphics[width=0.45\textwidth]{ 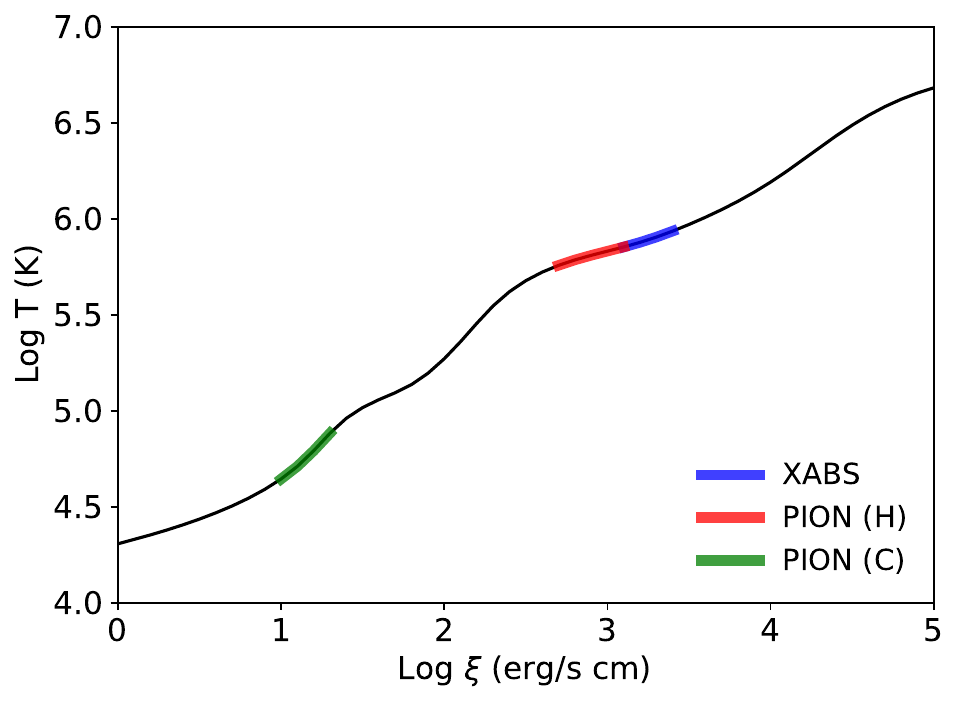}
		\includegraphics[width=0.45\textwidth]{ 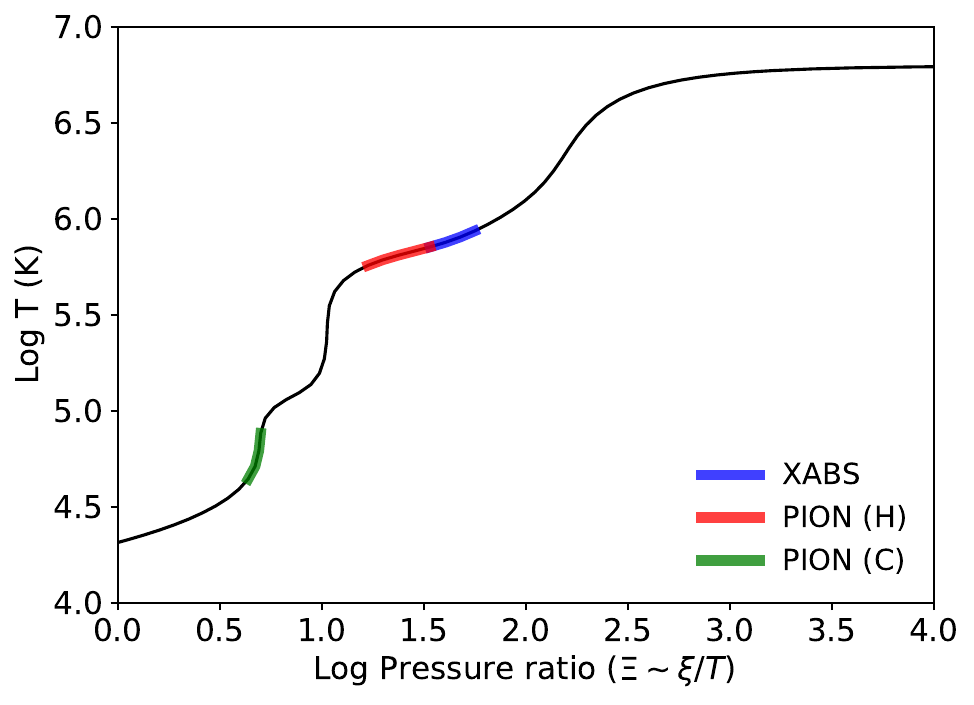}
        \vspace{-0.1cm}
		\caption{\small \textcolor{black}{{Spectral energy distribution (SED, top panel), photo-ionisation balance (middle panel), and log $T-\Xi$ stability curves (bottom panel). The plasma is expected to be generally stable}. Solid segments show the best-fit solutions for photoionised absorption (\textit{xabs}) and hot/cool emission (\textit{pion}) components.}}
		\label{fig: ionisation balance}
		\end{figure}

\textcolor{black}{Fig. \ref{fig: ionisation balance} shows the spectral energy distribution (SED, top panel), photo-ionisation balance (middle panel), and log $T-\Xi$ stability curves (bottom panel).} \\

\begin{figure}
		\centering
		\includegraphics[width=0.48\textwidth]{ 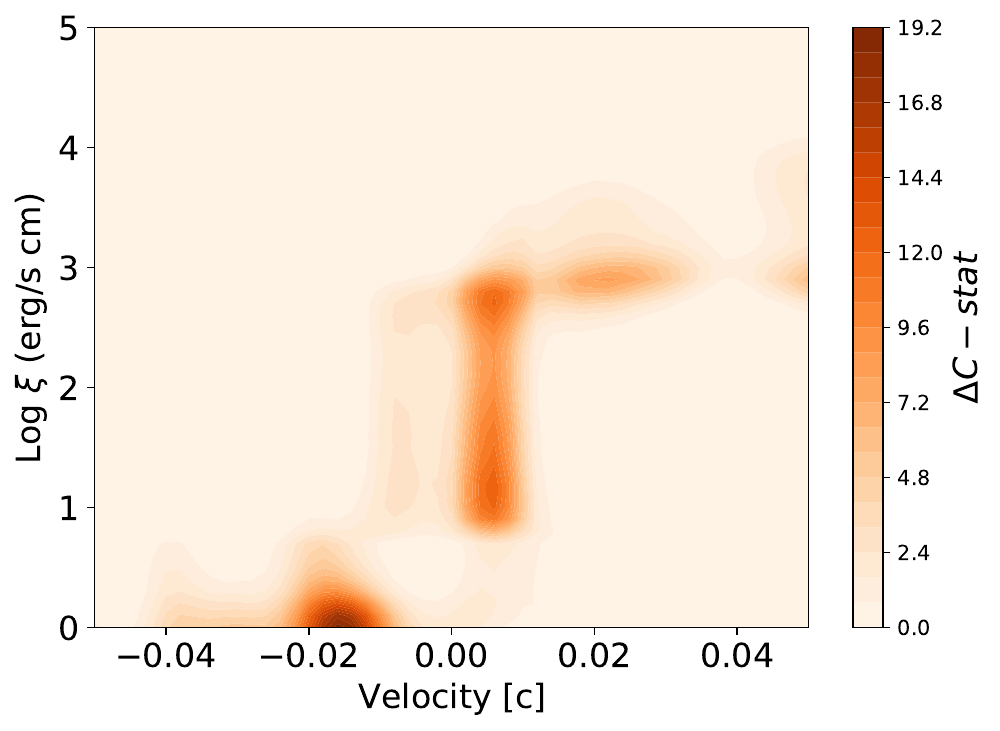}
        \includegraphics[width=0.48\textwidth]{ 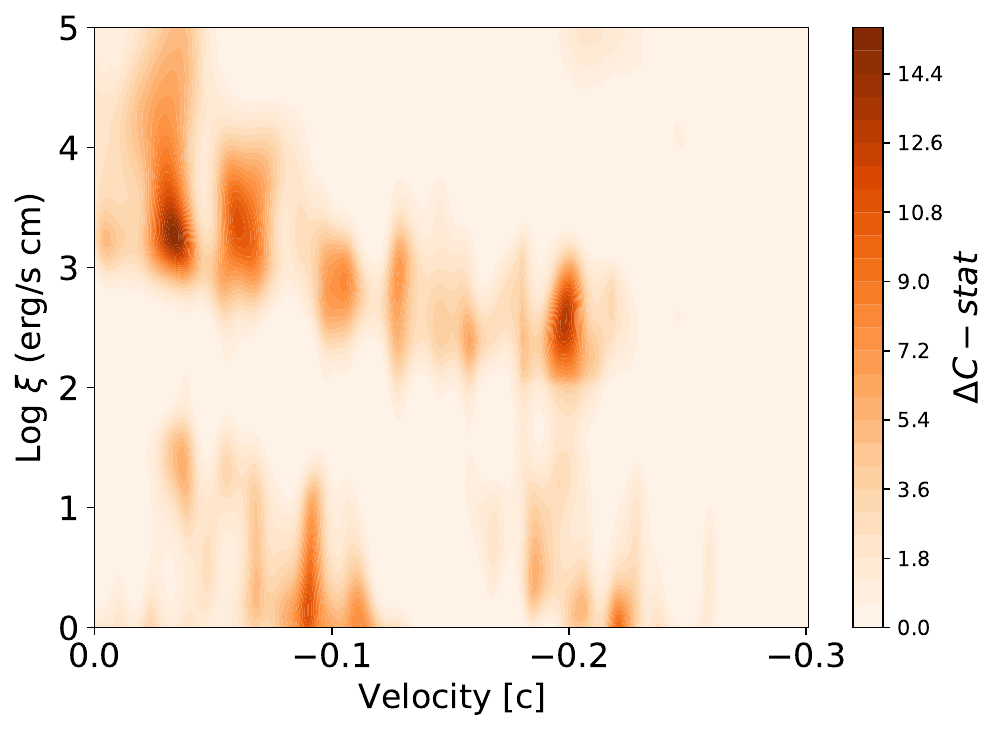}
		\caption{\small \textcolor{black}{{Top (bottom) panel: photoionised emission (absorption) model scan, right panel. These scans adopted a line-broadening of 1,000 km/s and Solar abundances. This of course decreased the chance of fake detections but on the other hand decreased the improvement to the spectral continuum as some elements, such as nitrogen, are clearly super-Solar with respect to e.g. oxygen (see Fig. \ref{fig: RGS spectra}).}}}
		\label{fig: physical model scan}
        \vspace{-0.3cm}
		\end{figure}

\textcolor{black}{Fig. \ref{fig: physical model scan} shows the photoionised emission (top panel) and absorption (bottom panel) model scan.}\\

\begin{figure}
		\centering
		\includegraphics[width=0.50\textwidth]{ 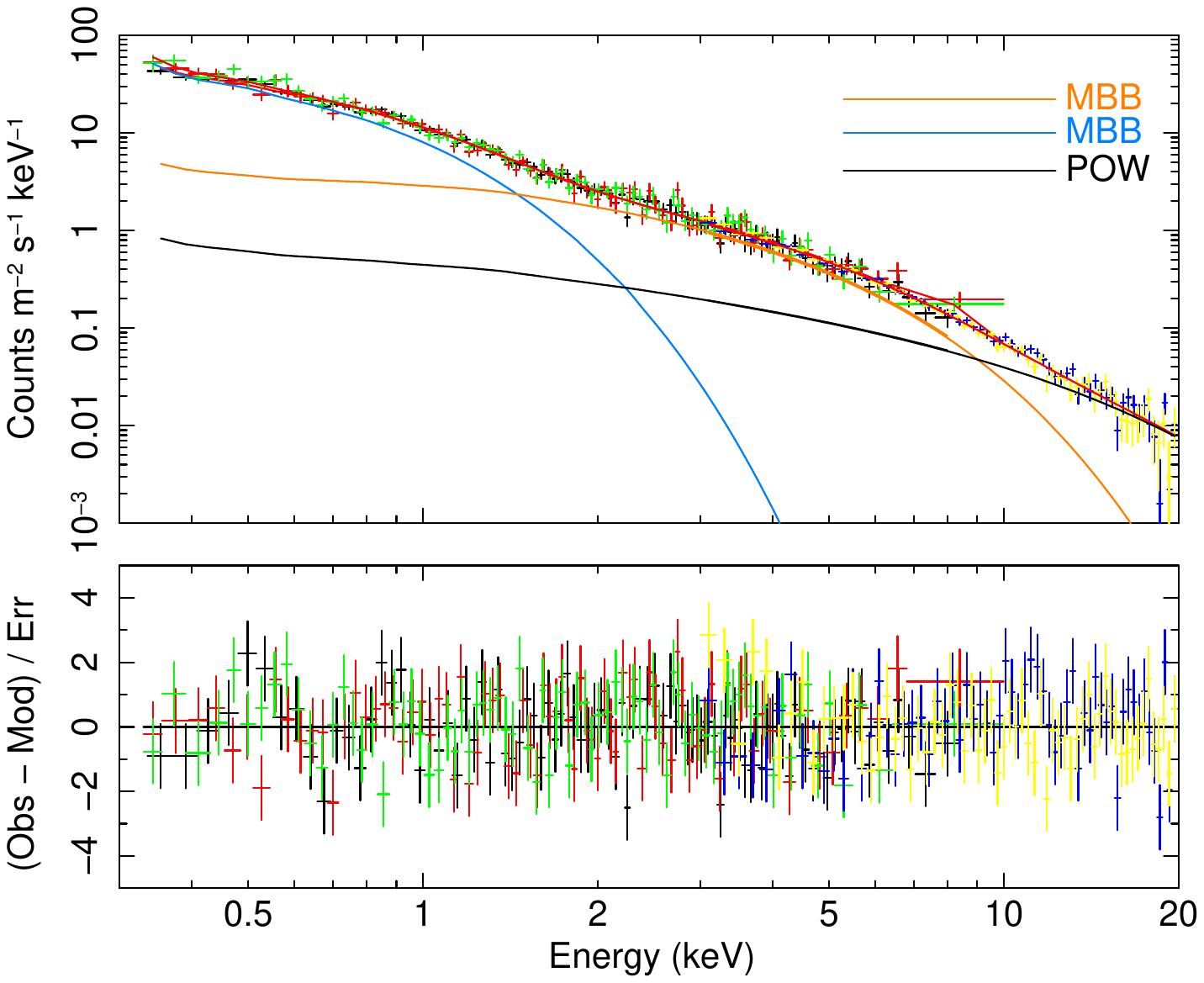}
        \vspace{-0.3cm}
		\caption{{\small XMM-Newton (obsid: 0724810301) + NuSTAR simultaneous data (2013). The spectra were fitted with a RHMM model plus a cutoff powerlaw. For the description of NuSTAR data is necessary the addition of a third component for the high-energy tail.}}
		\label{fig: XMM+NuSTAR modeling with RHMMP} 
		\end{figure}

\textcolor{black}{Fig. \ref{fig: XMM+NuSTAR modeling with RHMMP} shows the spectral modeling of the XMM-Newton \& NuSTAR simultaneous data with the RHMM model plus a cutoff powerlaw.  } \\ 

\begin{figure}
		\centering
        \includegraphics[width=0.48\textwidth]{ 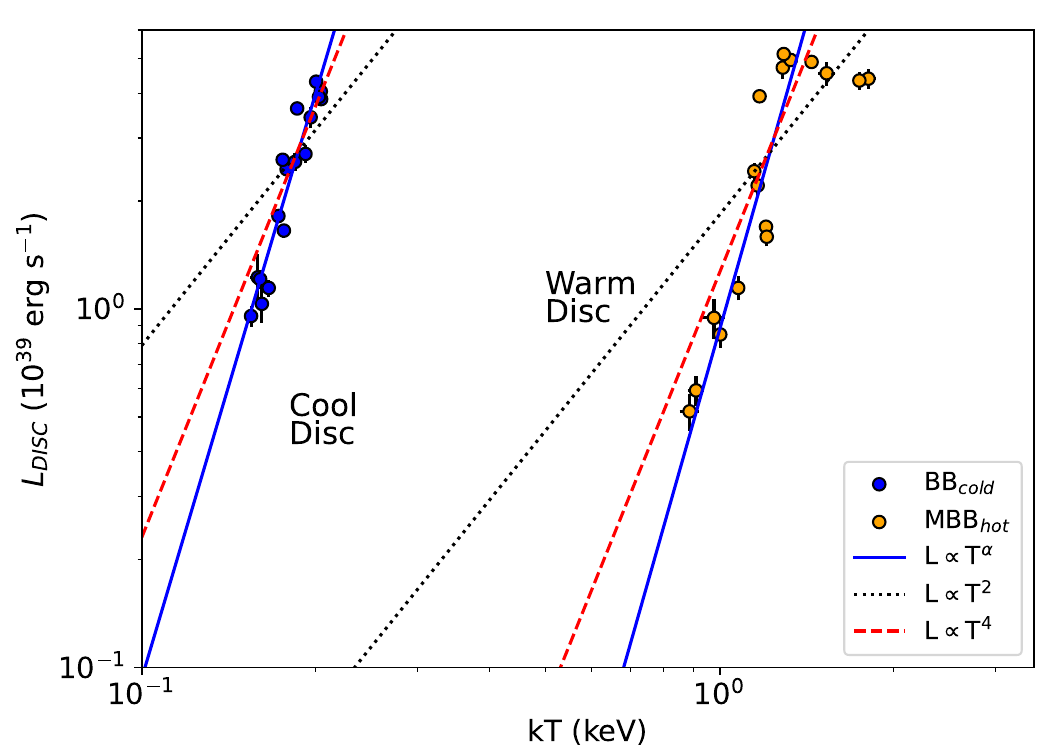}
        \includegraphics[width=0.48\textwidth]{ 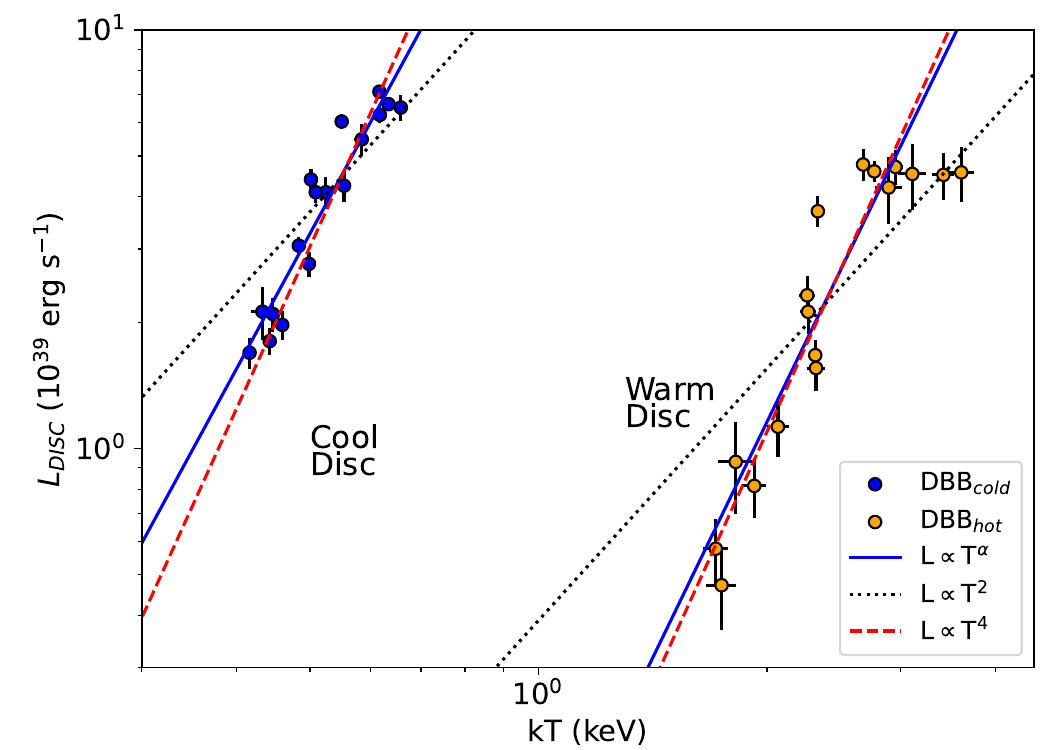}
        \includegraphics[width=0.46\textwidth]{ 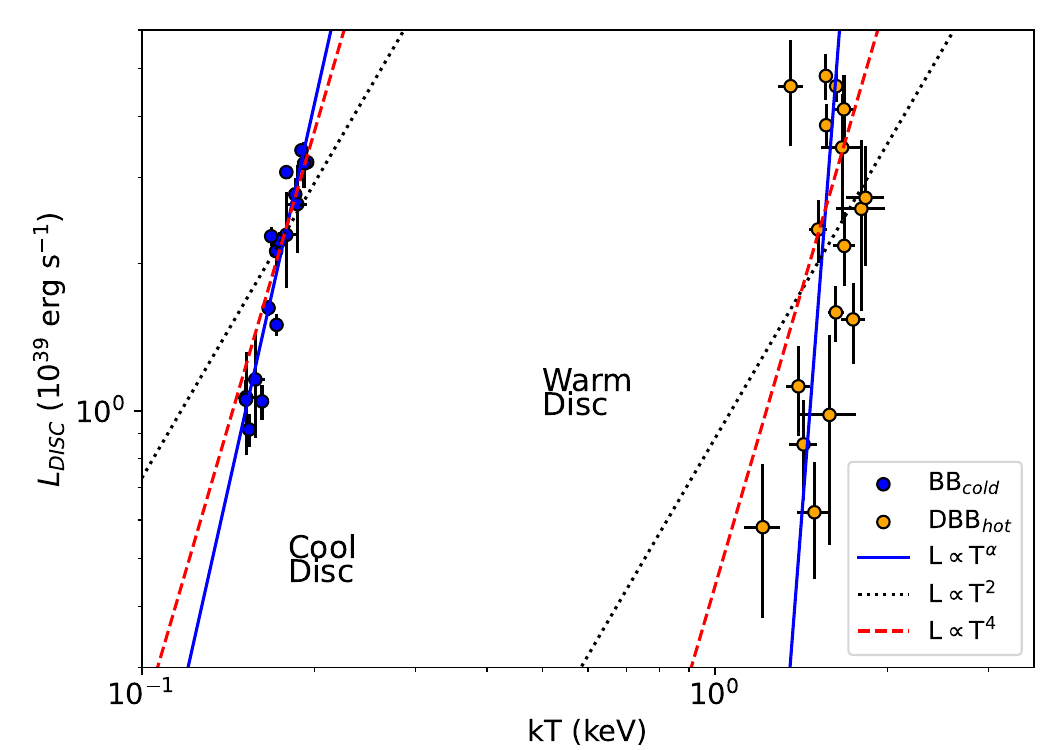}

		\caption{{\small L-T plots for the RHBM model (top panel), RHDD model (middle panel) and RHBDP model (bottom panel). Labels are the same as in Fig. \ref{fig: L-T plot}.}}
		\label{fig:L-T plots for alternative models}
        \vspace{-0.3cm}
		\end{figure}

\textcolor{black}{Fig. \ref{fig:L-T plots for alternative models} shows the luminosity-temperature trends with the following models: RHBM (top panel), RHDD model (middle panel) and RHBDP (bottom panel) model.  } \\

\textcolor{black}{Table \ref{table: Table results for the best fit RHBDP model for all observations. } reports the spectral fits results with RHBPD model for all observations. } \\

\textcolor{black}{Table \ref{table:pearsons and spearman correlation coefficients BM, POW} reports the slopes of the regression least square and the Pearson/Spearman coefficients for the L-T trends with the following models:  RHBM, RHBDP and RHDD model. } \\


\begin{center}
	\begin{table}
	\caption{Pearsons/Spearman coefficients and slope from the regression least square for the L-T trends, for both cool and hot components, by using the best fit values of the RHBM, RHBDP and RHDD  model.}  
	 \renewcommand{\arraystretch}{1.}
 \small\addtolength{\tabcolsep}{0pt}
 \vspace{0.3cm}
	\centering
	\scalebox{0.9}{%
  \begin{tabular}{c c c c c c c} 
    \hline
    \multirow{2}{*}{Correlation coefficient} &
      \multicolumn{2}{c}{RHBM}  & 
      \multicolumn{2}{c}{RHBDP} & 
      \multicolumn{2}{c}{RHDD} \\ \cline{2-7}
    &  ${(L-T)}_{cool}$ & ${(L-T)}_{hot}$ & ${(L-T)}_{cool}$ & ${(L-T)}_{hot}$ & ${(L-T)}_{cool}$ & ${(L-T)}_{hot}$  \\
    \hline
    Pearsons & 0.953 & 0.776 & 0.941 & 0.265 & 0.95 & 0.89  \\
    \hline
    Spearman & 0.928 & 0.836 & 0.924 & 0.306 & 0.94 & 0.87  \\
    \hline
    Slope & 5.42 $\pm$ 0.51 & 5.68 $\pm$ 1.22 & 5.22 $\pm$ 0.60 & 15.07 $\pm$ 6.39 & 3.33 $\pm$ 0.30 & 3.74 $\pm$ 0.51  \\
    \hline
  \end{tabular}}   \label{table:pearsons and spearman correlation coefficients BM, POW}
  \begin{quotation}
\end{quotation}
        \vspace{-0.3cm}
\end{table}
\end{center}

\label{lastpage}
\end{document}